%% file: main.tex
\documentclass[
  twocolumn,reprint,
  preprintnumbers,
  aps,
  prd, nofootinbib,
  superscriptaddress,
  floatfix
]{revtex4-2}

\usepackage{graphicx} 
\usepackage{dcolumn}  
\usepackage{bm}       
\usepackage{appendix}
\usepackage{xspace}
\usepackage[dvipsnames]{xcolor}
\usepackage{amsmath}
\usepackage{mathtools}
\usepackage{physics}
\usepackage[normalem]{ulem}
\usepackage{comment}
\usepackage[hidelinks]{hyperref}
\usepackage{cleveref}

\input{commands}

\begin{document}


\date{\today}
\preprint{CERN-TH-2026-213}
\title{Study of Radiative Muon Capture Experimental Data \texorpdfstring{\\}{ }
 and its Implications for Muon-Conversion Experiments}

 \author{Yury Kolomensky}
 \affiliation{Department of Physics, University of California, Berkeley, California 94720, USA}
 
\author{Michael MacKenzie}
\affiliation{Department of Physics, Yale University, New Haven, Connecticut 06520, USA}

\author{Ryan Plestid}
\affiliation{CERN, 1 Esplanade des Particules, CH-1211 Geneva 23, Switzerland}

\author{Matthew Stortini}
\affiliation{Department of Physics, Yale University, New Haven, Connecticut 06520, USA}

 \author{Ajla Trumic}
 \affiliation{Department of Physics, University of California, Berkeley, California 94720, USA}

\begin{abstract}
Radiative muon capture (RMC) is an important background in searches for the neutrinoless conversion of a muon into an electron in the field of a nucleus, and in particular in searches for the lepton flavor and lepton number violating process of muon-to-positron conversion.
The high-energy tail of the RMC photon spectrum remains theoretically understudied, leading to large uncertainties in background predictions for upcoming muon-conversion experiments.
In this work, we revisit previously published RMC data to investigate improved modeling of the high-energy tail.
We develop a model characterized by distinct nucleon-knockout final states, each with its associated multibody phase space and kinematic endpoint. This framework is applied to photon data from TRIUMF and positron data from SINDRUM~II, and the results are compared to the previously used closure approximation.
We find that separately modeling the distinct nucleon-knockout final states provides an improved description of the available RMC data near the endpoint, and we discuss the implications for the COMET and Mu2e experiments.
\end{abstract}

\maketitle

\input{introduction}

\input{theory}

\input{data}

\input{analysis}

\input{mu2e}

\section{Conclusion}

RMC is a critical background for the planned muon-conversion searches at COMET and Mu2e, particularly in the \mumep channel.
Theoretical modeling of the high-energy RMC photon spectrum remains limited, and commonly used approximations can systematically underestimate the endpoint region.
We find that the phase-space approximation introduced here better describes the high-energy tail of the RMC photon spectrum than the closure approximation, while enforcing fixed, physically motivated kinematic endpoint energies.
This makes it more suitable for estimating RMC backgrounds in upcoming muon-conversion searches.
The phase-space model also provides a physically motivated shape for the RMC contribution to the electron spectrum near the DIO endpoint. This may help constrain RMC contamination of the DIO tail (a dominant background for the $\mu^-\rightarrow {\rm e}^-$ search) in future measurements, alongside data-driven estimates from photon and positron spectra.

The measurements by the TRIUMF RMC Spectrometer Group \cite{RMC_1992_PhysRevC.46.1094, RMC_1998_PhysRevC.58.1767, RMC_1999_PhysRevC.59.2853} provide the highest-statistics RMC photon datasets available, covering a wide range of nuclear targets.
We find that the phase-space approximation describes the measured photon spectra well, consistently providing a better description than the closure approximation.
This work yields the first data-constrained nonzero estimate of the RMC photon rate within the final $\sim$10\,MeV below the kinematic endpoint for these nuclear targets.

The SINDRUM~II positron measurements provide additional datasets for evaluating RMC models, and notably include contributions from internal photon conversion.
We demonstrate that the phase-space approximation describes the observed SINDRUM~II high-energy positron spectrum tails well, improving the modeling of the SINDRUM~II gold target dataset and resolving the tension between the TRIUMF and SINDRUM~II preferred closure approximation endpoints.
These results indicate that the phase-space model provides significantly improved estimates of the RMC background for upcoming muon-conversion experiments.

The RMC background at muon-conversion experiments depends strongly on both the shape of the photon energy spectrum and the choice of nuclear target.
The upcoming COMET and Mu2e experiments will use aluminum stopping targets, for which the \mumep conversion energy lies well below the RMC kinematic endpoint.
The phase-space model fits presented here provide an initial estimate of the RMC background rate and shape that can guide analysis strategies for these next-generation experiments.
These results can also provide guidance for nuclear-target selection in muon-conversion searches beyond COMET and Mu2e.

\bigskip 

\begin{acknowledgments}
We thank Pavel Murat and Eleonora Diociaiuti for their contributions to initial studies of the published TRIUMF and SINDRUM~II data. Plestid thanks Richard Hill and Pavel Murat for many useful conversations.
MacKenzie and Stortini thank Sarah Demers for mentorship throughout this work.
The work by MacKenzie and Stortini were supported through award number DE-SC0017660.
This work was supported by the US Department of Energy (DOE)
Office of High Energy and Nuclear Physics under Contract Nos. DE-SC0017660 and DE-SC0018988, and by the Physics Department at the University of California, Berkeley. 
\end{acknowledgments}

\appendix
\section{Unfolded Data Values}
\label{app:unfolded_data}
To facilitate the use of the unfolded spectrum as an alternative model to phase-space fits for future phenomenological studies, the bin-by-bin values of the unfolded $^{27}\text{Al}$ dataset from Ref.~\cite{RMC_1992_PhysRevC.46.1094} are provided in Table~\ref{tab:unfolded_spectrum_full}.

\begin{table}[hbt!]
\centering
\caption{Unfolded $^{27}$Al spectrum data from Ref.~\cite{RMC_1992_PhysRevC.46.1094}. The energy bins use the bin center values and the unfolded counts and uncertainties are both given.}
\label{tab:unfolded_spectrum_full}
\begin{tabular}{ccc|ccc}
\hline\hline\\[-9pt]
$E$ (MeV) & Counts & Unc. & $E$ (MeV) & Counts & Unc. \\[1pt]
\hline
57.5 & 1658.0 & 1807.9 & 81.5 & 317.4 & 39.3 \\
58.5 & 1689.9 & 1589.3 & 82.5 & 261.0 & 34.2 \\
59.5 & 1800.5 & 1211.1 & 83.5 & 203.6 & 29.7 \\
60.5 & 2040.9 & 778.5 & 84.5 & 150.5 & 25.6 \\
61.5 & 2396.3 & 469.7 & 85.5 & 107.8 & 22.0 \\
62.5 & 2712.0 & 360.2 & 86.5 & 77.1 & 18.7 \\
63.5 & 2800.8 & 329.0 & 87.5 & 56.1 & 15.8 \\
64.5 & 2633.5 & 290.5 & 88.5 & 41.6 & 13.4 \\
65.5 & 2349.9 & 251.8 & 89.5 & 31.9 & 11.3 \\
66.5 & 2118.5 & 222.9 & 90.5 & 25.8 & 9.5 \\
67.5 & 1973.2 & 201.3 & 91.5 & 21.8 & 8.0 \\
68.5 & 1862.9 & 180.8 & 92.5 & 17.9 & 6.7 \\
69.5 & 1737.9 & 161.3 & 93.5 & 13.2 & 5.6 \\
70.5 & 1580.7 & 144.9 & 94.5 & 8.3 & 4.7 \\
71.5 & 1399.2 & 130.8 & 95.5 & 4.5 & 3.9 \\
72.5 & 1210.5 & 118.0 & 96.5 & 2.4 & 3.1 \\
73.5 & 1038.0 & 105.7 & 97.5 & 1.6 & 2.4 \\
74.5 & 900.7 & 93.9 & 98.5 & 1.6 & 1.9 \\
75.5 & 797.7 & 83.4 & 99.5 & 2.3 & 1.4 \\
76.5 & 708.5 & 74.2 & 100.5 & 3.2 & 1.1 \\
77.5 & 614.6 & 66.0 & 101.5 & 3.8 & 0.8 \\
78.5 & 520.5 & 58.4 & 102.5 & 3.5 & 0.6 \\
79.5 & 438.4 & 51.4 & 103.5 & 2.6 & 0.5 \\
80.5 & 373.7 & 45.0 & 104.5 & 1.3 & 0.7 \\
\hline\hline
\end{tabular}
\end{table}

\bibliographystyle{apsrev4-2}
\bibliography{bib/radiative_muon_capture,bib/clfv_lnv_refs,bib/bayesian_unfolding,bib/used_before, bib/tools}

\end{document}

%% file: commands.tex
\newcommand{\kmax}         {\mbox{$k_{\rm max}$}\xspace}
\newcommand{\kmaxstar}     {\mbox{$k_{\rm max}^{*}$}\xspace}
\newcommand{\kmaxstarstar}     {\mbox{$k_{\rm max}^{**}$}\xspace}

\newcommand{\mumem}        {\ensuremath{\mu^{-}\rightarrow \text{e}^{-}}\xspace}
\newcommand{\mumep}        {\ensuremath{\mu^{-}\rightarrow \text{e}^{+}}\xspace}

\newcommand{\zeronubb}     {\ensuremath{0\nu\beta\beta}\xspace}
\newcommand{\rmue}         {\ensuremath{\text{R}_{\mu \text{e}}}\xspace}
\newcommand{\MeVc}         {\text{MeV}\!/\!\ensuremath{c}\xspace}

\newcommand{\br}        {\ensuremath{\mathcal{B}}\xspace}

%% file: introduction.tex
\section{Introduction}

Charged lepton flavor violation (CLFV) is highly suppressed in the Standard Model with massive neutrinos.
The observation of a CLFV process would provide unambiguous evidence of physics beyond the Standard Model.
Potentially observable CLFV rates arise naturally in many new physics models, motivating a comprehensive experimental search program for such processes~\cite{2008_CLFV_Marciano,2018_CALIBBI_SIGNORELLI}.
Among these efforts, experiments using intense muon beams provide particularly sensitive probes, with next-generation searches expected to improve existing sensitivities by several orders of magnitude~\cite{2013_CLFV_MU3E_PROPOSAL, Bernstein_2019, 2020_COMET_Design_Report}.

This paper is motivated by the upcoming muonic-atom experiments COMET and Mu2e~\cite{bartoszek2015mu2etechnicaldesignreport, 2020_COMET_Design_Report}. In these experiments, negatively charged muons are guided onto a target where they are stopped and rapidly fall into atomic orbitals.
These experiments primarily search for coherent neutrinoless muon-to-electron conversion, \mumem, which produces a high-energy monoenergetic electron.
In addition, COMET and Mu2e are sensitive to neutrinoless muon-to-positron conversion, \mumep, a lepton flavor and number violating (LNV) process.
This process proceeds through an off-diagonal flavor transition that is complementary to the flavor-diagonal transitions probed by \zeronubb searches.

Standard Model processes in muonic atoms can mimic CLFV conversion signatures and therefore require precise modeling.
These processes include muon decay in orbit (DIO) and nuclear muon capture. 
Nuclear muon capture proceeds predominantly through ordinary muon capture (OMC), in which the bound muon is absorbed by the nucleus and a muon neutrino is emitted.
At the nucleon level, the charged-current interaction converts a proton into a neutron, with nuclear final states that may also include one or more emitted nucleons.
More rarely, the capture process is accompanied by the emission of a real or virtual photon; this process is known as radiative muon capture (RMC).
Real RMC photons can undergo Compton scattering or $e^+e^-$ pair production in the detector material, while virtual photons convert directly via internal pair creation.
The resulting high-energy electrons or positrons can subsequently be misidentified as a muon conversion candidate.

The COMET and Mu2e experiments utilize aluminum targets to search for both \mumem and \mumep conversion.
For aluminum, the kinematic endpoint of the RMC photon spectrum lies below the \mumem conversion energy, but above the endpoint for \mumep conversion.
While RMC is kinematically separated from \mumem searches, experimental sensitivity to \mumep conversion is highly dependent on the exact behavior of the RMC spectrum near its endpoint.
In contrast to the well-studied DIO spectrum~\cite{2012_CLFV_Czarnecki,2017_CLFV_Szafron_CE_LL}, the high-energy tail of the RMC spectrum remains theoretically understudied, leading to large uncertainties in background predictions.
To better understand this background, we reanalyze existing RMC data collected at TRIUMF \cite{RMC_1992_PhysRevC.46.1094} and SINDRUM~II \cite{SINDRUMII:1998mwd,2006_CLFV_SINDRUM_II_GOLD}.

High-statistics RMC measurements have historically modeled the spectral shape with the ``closure approximation''~\cite{Christillin:1979fa}. This shape is controlled by one parameter: the effective spectrum endpoint.
The endpoint is not predicted by the model and is instead extracted from the data. Fitted values of this effective endpoint typically lie well below the kinematic limit, leading to a systematic underestimation of the backgrounds for muon experiments.

In this work, we put forward an alternative treatment that better captures the structure of the RMC photon spectrum near the endpoint.
The spectrum is separated into contributions from distinct nucleon-knockout final states, each with its own physical kinematic endpoint.
For each final state, we neglect the photon-energy dependence of the squared matrix element, $|\mathcal{M}|^2$, resulting in a spectral shape that depends entirely on its multibody phase space.
The complete RMC spectrum is then modeled as a linear combination of these component spectra.
The individual branching fractions are treated as free parameters determined via direct fits to experimental data.
This approach provides a substantially better description of the available data than the closure approximation and yields more conservative background estimates for muon-conversion searches.

%% file: theory.tex
\section{Theoretical Overview}
\label{sec:rmc_overview}

Radiative muon capture is an important low-energy process that provides important insight into hadronic structure. Historically, RMC on a proton was ``prized'' for its sensitivity to the nucleon pseudoscalar form factor. In early literature, RMC was modeled using form factors~\cite{Primakoff:1959fs,Fujii:1959aa,Rood:1965aa,Christillin:1979fa,Christillin:1979pz,Christillin:1980nh,Fearing:1991je} with subsequent calculations being carried out in heavy-baryon chiral perturbation theory~\cite{Meissner:1997mq,Ando:1997es,Bernard:2000et}. To date, no application of chiral perturbation theory to nuclear RMC has been carried out.

Much of the literature on nuclear RMC has focused on the ``bulk'' of the photon spectrum. As such, nuclear details are often averaged over
and simplified treatments such as Fermi-gas models or the closure approximation supply reasonable descriptions of the data.
The situation at muon-conversion experiments is qualitatively different. Here, one is interested specifically in how to reliably extrapolate the spectrum into the endpoint region. The endpoint region is sensitive to details of the kinematics such as the energy available in the decay. These details are not well captured by simplified nuclear treatments such as the Fermi gas model~\cite{Christillin:1979fa,Fearing:1991je} or the closure approximation~\cite{Primakoff:1959fs}, which are instead designed to capture the integrated strength of many nuclear final states rather than a select few allowed by kinematics. 

In what follows, we first describe the general theory of RMC. Next, we describe the traditionally used closure approximation. This was the method used to analyze TRIUMF data, which results in an uncontrolled extrapolation to the endpoint. Finally, we outline a different strategy in which the energy dependence of the  matrix element is neglected, and only the phase space and kinematic thresholds are included in the fit. This latter approach offers a better-controlled extrapolation to the endpoint, and can be improved with nuclear modeling of the relevant nuclear matrix element.  

\subsection{Kinematics and Phase Space }
\label{sec:kinematics}
Let us consider the inclusive reaction 
\begin{equation}
    \label{RMC-kinematics-def}
    A(\vb{0}) \rightarrow X(-\vb{p}-\vb{k}) + \nu (\vb{p}) + \gamma(\vb{k})~,
\end{equation}
where $A$ is a muonic atom and $X$ is any hadronic state with center-of-mass momentum $-\vb{p}-\vb{k}$. We may write the inclusive decay rate as a sum of exclusive modes, 
\begin{equation}
    \frac{\dd \Gamma}{\dd k} = \qty[\frac{\dd \Gamma}{\dd k}]_{0N} + \qty[\frac{\dd \Gamma}{\dd k}]_{1N} + \qty[\frac{\dd \Gamma}{\dd k}]_{2N}  + \ldots~,
\end{equation}
where the subscript denotes the number of knocked-out nucleons. For example, $0N$ corresponds to an intact nucleus in the final state. Within each of these categories one should also sum over excited states of the nuclear system, i.e., 
\begin{equation}
    \qty[\frac{\dd \Gamma}{\dd k}]_{0N} = \sum_n  \qty[\frac{\dd \Gamma^{(n)}}{\dd k}]_{0N}~, 
\end{equation}
with $(n)$ denoting the state of the nucleus. 

Notice that the ground-state term with no nucleon knockout $[\dd \Gamma^{(0)}/\dd k]_{0N}$ will generally have the highest endpoint energy. Other final states have systematically \textit{lower} endpoints (cf.\ Table~\ref{tab:knockout_endpoint_energies}). Moreover, the phase-space measure is different depending on the multiplicity of the final-state particles.

Let us make this concrete. Consider the $0N$ decay rate. The Lorentz-invariant phase space is given by (see, e.g., Ref.~\cite{Fearing:1979be})
\begin{equation}
    \begin{split}
    \dd\Phi_{3}= \qty[\frac{\dd^3k}{(2\pi)^3 2 |\vb{k}|}]\qty[\frac{\dd^3p}{(2\pi)^3 2 |\vb{p}|}]&\qty[\frac{\dd^3p_B}{(2\pi)^3 2 E_B}]\\
    &\times (2\pi)^4~\delta^{(4)}(\Sigma P_3)~,
    \end{split}
\end{equation}
where $\Sigma P_3 = p+k+p_B-p_A$ is the four-momentum conservation condition for three-body phase space. The differential decay rate is given by 
\begin{equation}
   \qty[ \frac{\dd \Gamma^{(0)}}{\dd k}]_{0N} = \frac{1}{2 M_A} \int \dd\Phi_{3} \langle |\mathcal{M}|^2 \rangle \delta(\omega-k) ~.
\end{equation}
Evaluating the phase-space integral in the limit where $M_A \sim M_B \gg m_\mu$ leads to a spectrum of the form~\cite{Fearing:1979be,PlestidAndHill}
\begin{equation}
    \qty[ \frac{\dd \Gamma^{(0)}}{\dd k}]_{0N} \propto  \qty[\frac{\langle |\mathcal{M}|^2 \rangle }{2E_\nu} ] \times  k E_\nu^2~,
\end{equation}
where the angle brackets now include averages over spin and angles, and
\begin{equation}
    E_\nu = (k_{\rm max,0} - k)~,
\end{equation}
with $k_{\rm max,0}$ the endpoint energy for 0-nucleon knockout. We will discuss different \textit{ans\"atze} for $\langle |\mathcal{M}|^2 \rangle /2E_\nu$ in what follows. 

Let us next consider single nucleon-knockout. We now have four-body phase space 
\begin{equation}
    \begin{split}
    \dd\Phi_{4}= &\qty[\frac{\dd^3k}{(2\pi)^3 2 |\vb{k}|}]\qty[\frac{\dd^3p}{(2\pi)^3 2 |\vb{p}|}]\qty[\frac{\dd^3p_B}{(2\pi)^3 2 E_B}]\\
    &~~~~~~~~~\qty[\frac{\dd^3p_N}{(2\pi)^3 2m_N}]\times (2\pi)\!\!~^4~\delta^{(4)}(\Sigma P_4)~.
    \end{split}
\end{equation}
The neutrino energy is now given by 
\begin{equation}
    E_\nu = (k_{\rm max,1} - k-T_N)~,
\end{equation}
where $k_{\rm max,1}$ is the endpoint for one-nucleon knockout and $T_N$ is the kinetic energy of the knocked out nucleon. One must average over phase space of the knocked out nucleon  $\int_0^{p_{\rm max}} \dd p_N p_N^2(k_{\rm max,1}-p_N^2/2m_N - k)^2$. This modifies the shape of the photon spectrum, especially as $k\rightarrow k_{\rm max,1}$.

More generally, the angle-averaged matrix element will have some non-trivial dependence on $k$. In what follows we do not attempt a detailed modeling of this dependence. We now turn to a discussion of the closure approximation and then to a ``phase-space-only'' analysis strategy.

\subsection{Closure Approximation}
\label{sec:closure_approx}

The RMC spectrum has historically been modeled using the closure approximation, in which the individual nuclear final states are replaced by a single effective transition~\cite{Christillin:1979fa}. This amounts to the \textit{ansatz}
\begin{equation}
    \qty[\frac{\langle |\mathcal{M}(k)|^2 \rangle }{2E_\nu} ] \propto (1-2x+2x^2)\Theta(1-x)\,,
\end{equation}
where $x=k/\kmax$, $\Theta$ is a Heaviside function, and \kmax is treated here as a free parameter (i.e.\ not fixed by any physical kinematic thresholds). The resulting approximate photon spectrum is given by
\begin{equation}
    R(x) = A (1-2x+2x^2) x (1-x)^2 \Theta(1-x)\,,
    \label{eq:closure}
\end{equation}
where $A$ is the normalization, $x = k/\kmax$, and \kmax is the endpoint associated with this effective transition.

The closure approximation can describe the total RMC rate and the bulk of the photon spectrum, but it was not intended to precisely model the high-energy tail.
Because \kmax does not necessarily coincide with the true kinematic limit, the model predicts no photons above the fitted endpoint even when such photons are kinematically allowed.
Since photons in this region can contribute to backgrounds in muon-conversion searches, this artificial truncation may lead to a significant underestimate of the RMC background in high-sensitivity \mumep searches.

\subsection{Phase-Space Approximation}
\label{sec:plestid_approx}

Here we pursue an alternative method to estimate the RMC photon spectrum.
We ignore the photon energy dependence of the RMC matrix element, assuming $\langle |\mathcal{M}|^2\rangle /(2E_\nu)$ is constant. This leaves only the phase-space measure to sculpt the resulting photon spectrum.
The photon spectrum is then approximated as the sum of the exclusive channels with undetermined branching ratios but \textit{fixed and physical} endpoint energies. 

In what follows we will use the \textit{ansatz} 
\begin{equation}
    \label{phase-space-only-ansatz}
    \frac{\langle |\mathcal{M}|^2 \rangle}{2E_\nu} = {\rm constant}~. 
\end{equation}
The matrix element squared will always be proportional to $2E_\nu$ because of the relativistic normalization of states, and it is therefore convenient to work with $|\mathcal{M}|^2/(2E_\nu)$.

Let us consider the behavior of the 0-nucleon knockout spectrum, 
\begin{equation}
    \qty[ \frac{\dd \Gamma^{(0)}}{\dd k}]_{0N} \propto  \qty(1- x_0)^2 x_0\Theta(1-x_0) ~,
\end{equation}
for  $x_0= k/k_{\rm max,0}$. This equation shows how the phase space closes off like $E_\nu^2$ as $E_\nu\rightarrow 0$, which defines the endpoint region.  The generalization to include excited final states of the nucleus is straightforward: one simply adjusts the endpoint energy.  

Next, consider $1N$ knockout. The neutrino energy is now given by 
\begin{equation}
    E_\nu = (k_{\rm max,1} - k-T_N)~,
\end{equation}
with $T_N$ the nucleon kinetic energy and $k_{\text{max},(n=1,1N)}$ accounting for the binding energy of the nucleon. Using the non-relativistic approximation to $T_N$, using the \textit{ansatz} described above for the matrix element, and carrying out the $p_N$ phase-space integral we encounter $\int_0^{p_{\rm max}} \dd p_N p_N^2(k_{\rm max,1}-p_N^2/2m_N - k)^2 \sim (k_{\rm max,1}-k)^{7/2}$ where $p_{\rm max} = \sqrt{2m_N(k_{\rm max,1}-k)}$.
We thus find 
\begin{equation}
    \qty[ \frac{\dd \Gamma^{(0)}}{\dd k}]_{1N} \propto \qty(1- x_1)^{7/2} x_1 \Theta(1-x_1)~,
\end{equation}
where $x_1=k/k_{\rm max,1}$. Again, the generalization to include excited final states of the nucleus is straightforward: for $N$-nucleon knockout one finds a shifted endpoint with a spectrum that goes to zero like $(1-x_N)^{2 + 3N/2}$.

We now define the ``phase-space'' approximation using the \textit{ansatz} given in Eq.~\ref{phase-space-only-ansatz}. For simplicity, all nuclear excited states are treated as having the same endpoint energy, and the results are summed together. The assumed spectral shapes for i-nucleon-knockout RMC are given by
\begin{equation}
    \qty[\frac{\dd \Gamma}{\dd k}]_{iN}= A_i x_i(1 - x_i)^{2+3i/2} \Theta(1-x_i)\,,
    \label{eq:plestid_0n}
\end{equation}
where $x_i = k/k_{\text{max},i}$ and $A_i$ is a floating normalization. The endpoint energies $k_{\text{max},i}$ are determined by computing the neutron or proton removal energies. This approximation amounts to ignoring all energy dependence of the nuclear matrix element \textit{except} for the trivial $2E_\nu$ factor related to the relativistic normalization of states. 

This approach does not account for the distribution of strength and endpoint energy among nuclear final states. It also neglects photon-energy dependence in the nuclear matrix element. There are therefore multiple improvements that could be made in the future. However, as we discuss below, this simple approach already significantly improves the description of the TRIUMF and SINDRUM~II data relative to the closure approximation. Since $k_{\text{max},i}$ are fixed and physical, this treatment also provides a more reliable extrapolation of the spectrum into the near-endpoint region relevant for muon-conversion experiments. 

%% file: data.tex
\section{Existing Experimental Data}
\label{sec:existing_data}

Direct measurements of the RMC photon spectrum are limited, particularly in the high-energy region relevant to future muon-conversion experiments.
The highest-statistics measurements were performed by the TRIUMF RMC Spectrometer Group in the 1990s.
The SINDRUM~II muon-conversion experiment, though not dedicated to RMC, measured inclusive electron and positron datasets from stopped muons; these spectra naturally include contributions from RMC.

\subsection{TRIUMF Measurements}

The TRIUMF RMC measurements spanned several nuclear targets, representing the largest and most diverse RMC data catalog available.
These spectra were reconstructed through $e^+e^-$ pair conversions using a lead converter and a tracking spectrometer~\cite{RMC_1992_WRIGHT_1992_249}.
To avoid the overwhelming background from DIO electrons, the photon spectra were measured above $57\,\text{MeV}$.

The measured spectra were fitted using the closure approximation, with the normalization and spectral endpoint treated as free parameters.
Each dataset contains a few thousand reconstructed photons with measured RMC photon rates above $57\,\text{MeV}$ of $\mathcal{O}(10^{-5})$ with respect to OMC~\cite{RMC_1991_PhysRevC.43.1425, RMC_1992_PhysRevC.46.1094, RMC_1998_PhysRevC.58.1767, RMC_1999_PhysRevC.59.2853}.
Across these measurements, the fitted closure endpoint was typically $\sim\!\!10\,\text{MeV}$ below the kinematic limit for the corresponding nuclear target.

\subsection{SINDRUM~II Measurements}

The SINDRUM~II experiment stopped significantly more muons on a nuclear target than the TRIUMF RMC measurements \cite{SINDRUMII:1998mwd,2006_CLFV_SINDRUM_II_GOLD}.
A cylindrical tracking detector surrounded a muon stopping target to reconstruct high-momentum charged particles.
Because the experiment focused primarily on muon-conversion searches, no dedicated photon pair-conversion dataset was published.
However, high-momentum electron and positron spectra were published using titanium and gold targets.
While the electron datasets are dominated by DIO electrons, the positron spectra provide clean channels primarily composed of asymmetric RMC photon conversions.
These datasets therefore provide one of the highest-statistics probes of the high-energy tail of the RMC spectrum.

A detailed analysis of the RMC spectrum was not performed by the SINDRUM~II collaboration, as the RMC background in the muon-conversion signal region was negligible \cite{SINDRUMII:1998mwd, 2006_CLFV_SINDRUM_II_GOLD}.
However, a comparison of the measured spectrum with the closure-approximation prediction revealed a significant excess in the high-energy tail~\cite{SINDRUMII:1998mwd, 1997_CLFV_SINDRUM_II_Kaulard_Thesis}.
This discrepancy suggests that the closure approximation breaks down at the sensitivity reached by SINDRUM~II.
As future muon-conversion experiments will accumulate approximately four orders of magnitude more muon captures, a more sophisticated description of the RMC background is necessary.

%% file: analysis.tex
\input{analysis_triumf}
\input{analysis_sindrum}

%% file: analysis_triumf.tex
\section{Analysis of TRIUMF Data}
\label{sec:analysis_triumf}

The TRIUMF RMC datasets span the 1990s~\cite{RMC_1991_PhysRevC.43.1425, RMC_1992_PhysRevC.46.1094, RMC_1995_Bergbusch_MS_thesis, RMC_1998_PhysRevC.58.1767, RMC_1999_PhysRevC.59.2853}.
We restrict our analysis to the 1992 datasets. The corresponding publication provides an analytic detector response model, allowing us to convolve theoretical spectra with the provided response to produce folded spectra. Subsequent publications instead rely on simulation-based detector descriptions for which no equivalent response is provided, preventing the detector convolution from being reproduced using the published information alone. Furthermore, closure approximation curves shown overlaid with later data differ markedly from those published in 1992 despite identical theoretical models, indicating that later datasets cannot be adequately described by the published response function. 

We first perform validation and consistency tests of the published response function and our convolution procedure.
For this purpose, we digitize the published radiative pion capture (RPC) calibration data
along with closure approximation curves that TRIUMF overlaid on the measured RMC spectra for several values of \kmax.
Next, we fit the published data for several nuclear targets using the closure approximation and compare our results with those reported by TRIUMF.
This same forward-convolution framework is subsequently used to fit the phase-space model introduced in Sec.~\ref{sec:plestid_approx} to the RMC data.
Finally, we present a complementary Bayesian unfolding analysis to reconstruct the underlying physical photon spectrum.

\subsection{Detector Response}
\label{sec:triumf_response}
The detector response parameterization published in Ref.~\cite{RMC_1992_PhysRevC.46.1094} is expressed as a function of the true photon energy, $E_\gamma$, and the reconstructed energy, $E$. 
We convolve theoretical RMC spectra with this response and fit the resulting reconstructed-energy distributions to the published data.
The response is given by
\begin{subequations}
\label{eq:triumf_response}
\begin{align}
P(E_\gamma,E) ={}&
A \exp\left[
    -\frac{1}{2\sigma_0^2}(E-E_0)^2
\right]
\nonumber\\
&\qquad
+ F \exp\left[
    -\frac{1}{2\sigma_3^2}(E-E_3)^2
\right]
\nonumber\\[0.5em]
&\hspace{8em}
\text{for } E_1 < E < E_2,
\label{eq:triumf_response_middle}
\\[0.5em]
P(E_\gamma,E) ={}&
B \exp\left[
    -\frac{1}{\sigma_1}(E_1-E)
\right]
\nonumber\\
&\qquad
+ F \exp\left[
    -\frac{1}{2\sigma_3^2}(E-E_3)^2
\right]
\nonumber\\[0.5em]
&\hspace{10.25em}
\text{for } E \leq E_1,
\label{eq:triumf_response_low}
\\[0.5em]
P(E_\gamma,E) ={}&
C \exp\left[
    -\frac{1}{\sigma_2}(E-E_2)
\right]
\nonumber\\
&\qquad
+ F \exp\left[
    -\frac{1}{2\sigma_3^2}(E-E_3)^2
\right]
\nonumber\\[0.5em]
&\hspace{10.5em}
\text{for } E \geq E_2,
\label{eq:triumf_response_high}
\end{align}
\end{subequations}
with 
\begin{equation}
\begin{aligned}
E_1 &= E_0 - \frac{\sigma_0^2}{\sigma_1},
&
E_2 &= E_0 + \frac{\sigma_0^2}{\sigma_2},
\\
B &= A\exp\left(-\frac{\sigma_0^2}{2\sigma_1^2}\right),
&
C &= A\exp\left(-\frac{\sigma_0^2}{2\sigma_2^2}\right).
\end{aligned}
\end{equation}
The response parameters are represented by polynomials in $E_\gamma$, and their corresponding coefficients are summarized in Table~\ref{tab:triumf_response_params}.
The additional Gaussian term with coefficient $F$ is only included in the response for $E_\gamma$ above $60\,\text{MeV}$.

\begin{table}[!h]
\caption{
  Polynomial parameterization coefficients (Par.) for the TRIUMF RMC spectrometer detector response as presented in Tables VII and VIII in Ref.~\cite{RMC_1992_PhysRevC.46.1094}.
  Each parameter is modeled as $P = c_0 + c_1 E_\gamma + c_2 E_\gamma^2 + c_3 E_\gamma^3$, where $E_\gamma$ is the true photon energy.
  The parameter $E_0^\text{L}$ ($E_0^\text{H}$) is used as $E_0$ in Eq.~(\ref{eq:triumf_response}) when $E_\gamma < 60$\,MeV ($E_\gamma \geq 60$\,MeV). The coefficients are given for $E_\gamma$ in units of MeV.
}
\centering
\renewcommand{\arraystretch}{1.3}
\begin{tabular}{lcccc}
\hline\hline
Par. & $c_0$ & $c_1$ & $c_2$ & $c_3$\\ \hline
$\sigma_0$ & -0.5836 & 0.0352 & - & -\\
$\sigma_1$ & -5.879 & 0.1653 & -5.149$\times 10^{-4}$ & -\\
$\sigma_2$ & 1.596 & -0.03859 & 3.883$\times 10^{-4}$ & -\\
$\sigma_3$ & -47.80 & 1.010 & -4.406$\times 10^{-3}$ & -\\
E$_3$ & 1.068 & 0.7507 & - & -\\
E$_0^\text{H}$ & -1.161 & 0.9481 & 1.724$\times 10^{-4}$ & - \\
E$_0^\text{L}$ & 22.73 & 0.1995 & 5.993$\times 10^{-3}$ & - \\
A & 3.259$\times 10^{-4}$ & -4.120$\times 10^{-4}$ & 1.015$\times 10^{-5}$ & -4.05$\times 10^{-8}$  \\
F/A & -0.1337 & 2.828$\times 10^{-3}$ & -9.701$\times 10^{-6}$ & \\
\hline\hline
\end{tabular}
\label{tab:triumf_response_params}

\end{table}

\subsection{Validating the Detector Response}
\label{sec:triumf_response_validation}

\begin{figure}
    \centering
    \includegraphics[width=0.99\linewidth, trim = 0 0 0 0, clip]{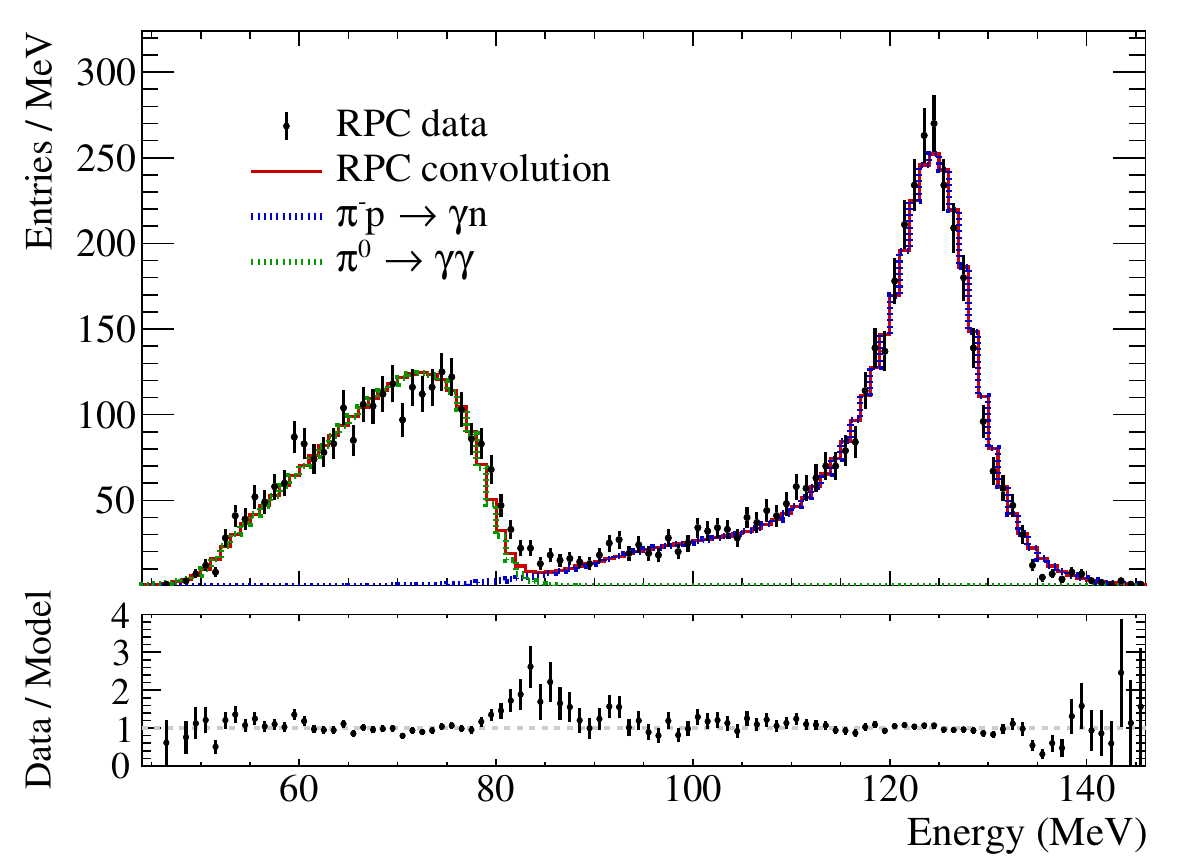}
    \caption{Digitized RPC calibration data from Ref.~\cite{RMC_1992_PhysRevC.46.1094} compared with the theoretical RPC spectrum convolved with the published detector response.}
    \label{fig:triumf_rpc_comparison}
\end{figure}

Several tests were performed to validate the published detector response function and verify our convolution procedure.
The first used RPC calibration data provided alongside the RMC measurements in Ref.~\cite{RMC_1992_PhysRevC.46.1094}.
The RPC calibration used a liquid-hydrogen target and reconstructed photons produced through either the monoenergetic channel $\pi^- p \rightarrow n\gamma$, or the charge-exchange channel $\pi^- p \rightarrow \pi^0 n$ followed by $\pi^0 \rightarrow \gamma\gamma$.
The former process produces $129.4\,\text{MeV}$ photons, while the latter yields a flat photon spectrum between $54.9$ and $83.0\,\text{MeV}$.
Figure~\ref{fig:triumf_rpc_comparison} compares the digitized calibration data with the theoretical RPC spectrum convolved with the published detector response.
The two theoretical components are normalized separately using the corresponding regions of the measured spectrum.
Overall, the convolved model describes the calibration data well, though it underestimates the counts in the high-energy tail of the $\pi^0\rightarrow\gamma\gamma$ data.

\begin{figure}
\centering
    \includegraphics[width=0.99\linewidth, trim = 0 0 0 0, clip]{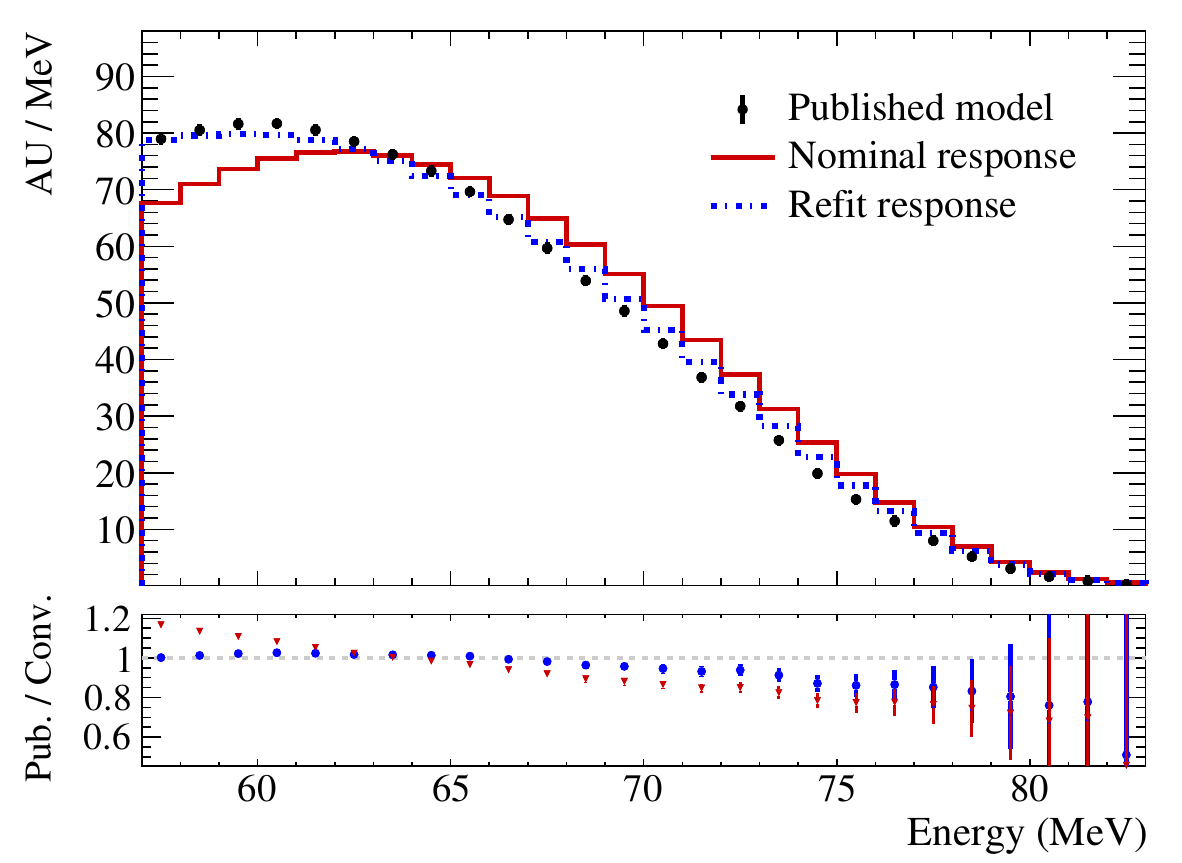}
    \caption{Digitized closure approximation curve with $\kmax=85\,\text{MeV}$ published in Ref.~\cite{RMC_1992_PhysRevC.46.1094} compared with convolutions using the published and refit detector responses.}
    \label{fig:triumf_mc_vs_convolutions_85}
\end{figure}

\begin{figure}
\centering
    \includegraphics[width=0.99\linewidth, trim = 0 0 0 0, clip]{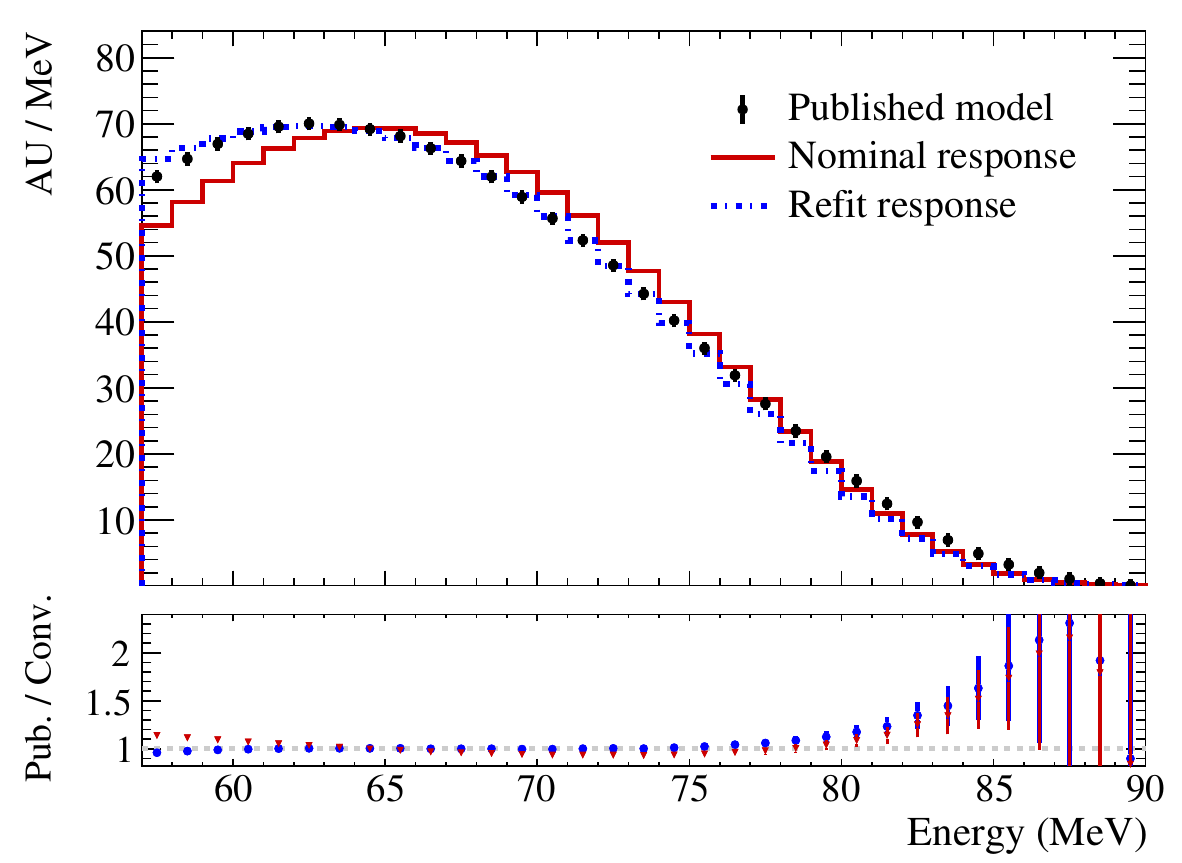}
    \caption{Digitized closure approximation curve with $\kmax=90\,\text{MeV}$ published in Ref.~\cite{RMC_1992_PhysRevC.46.1094} compared with convolutions using the published and refit detector responses.}
    \label{fig:triumf_mc_vs_convolutions_90}
\end{figure}

As a second test, we digitize the closure approximation curves overlaid on the measured RMC spectra for several values of \kmax in Ref.~\cite{RMC_1992_PhysRevC.46.1094}.
These curves are described as being produced by convolution of the theoretical spectra with the analytic detector response model.
We repeat this procedure and compare the results with the digitized curves.
The comparisons for $\kmax=85\,\text{MeV}$ and $\kmax=90\,\text{MeV}$ are shown in Figs.~\ref{fig:triumf_mc_vs_convolutions_85} and~\ref{fig:triumf_mc_vs_convolutions_90}, respectively.
In both cases, the agreement is poor.
The two curves also exhibit inconsistent behaviors, such as over- and under-estimated high-energy tails.

The discrepancy between our convolutions and the published closure approximation curves suggests that the curves may have been produced using a detector-response treatment that differs from the published analytic parameterization.
This interpretation is consistent with the transition to simulation-based detector models in later TRIUMF analyses.
The disagreement is largest at low reconstructed energies, while the monoenergetic RPC peak is reproduced well.

\begin{figure}
\centering
    \includegraphics[width=0.99\linewidth, trim = 0 0 0 0, clip]{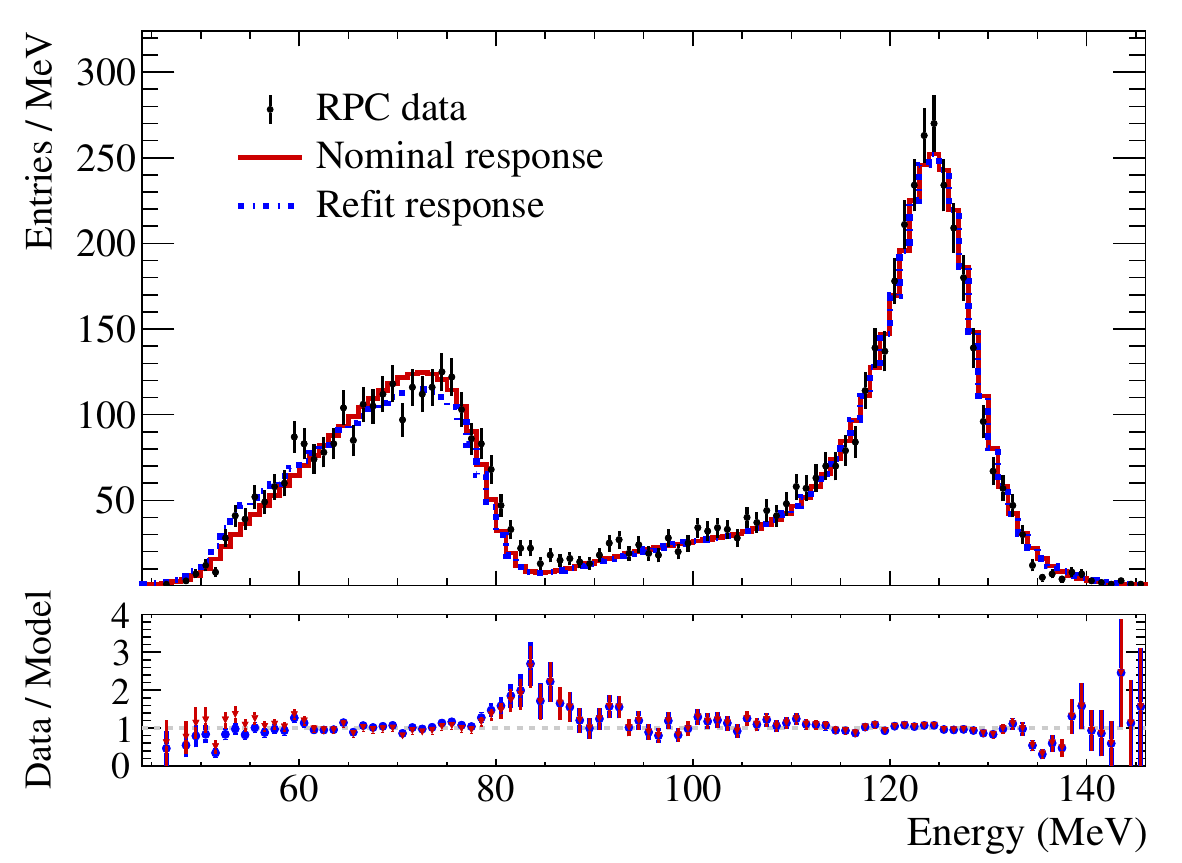}
    \caption{Digitized RPC calibration data from Ref.~\cite{RMC_1992_PhysRevC.46.1094} compared with theoretical RPC spectra convolved with the published detector response and the response obtained by refitting $A(E_\gamma)$.}
    \label{fig:triumf_rpc_model_comparison}
\end{figure}

In order to better reproduce the published closure approximation curves, we fit the polynomial coefficients for $A(E_\gamma)$ to the digitized closure approximation curves and RPC calibration data.
The resulting polynomial coefficients are listed in Table~\ref{tab:triumf_alt_response_params}.
As shown in Figs.~\ref{fig:triumf_mc_vs_convolutions_85}, \ref{fig:triumf_mc_vs_convolutions_90}, and~\ref{fig:triumf_rpc_model_comparison}, the refitted response provides a better description of the published closure approximation curves and the RPC data.
Modifying additional parameters of the response does not achieve a notably better agreement without introducing unphysical features into the response.

We carry three detector-response treatments into the subsequent analysis: the published response, serving as the nominal model; the alternative response, using the refitted $A(E_\gamma)$ polynomial coefficients; and the published response, evaluated with the fit restricted to a reconstructed $E>65\,\text{MeV}$.
This last treatment is to account for potential turn-on effects in the geometric and selection acceptance.
The envelope of the results obtained with these three treatments is used to estimate the systematic uncertainty associated with the modeling of the detector response.

\begin{table}[!h]
\caption{
  Refitted response function parameter polynomial coefficients, where the nominal parameters from Table~\ref{tab:triumf_response_params} are used for all other polynomial coefficients.
}
\label{tab:triumf_alt_response_params}
\centering
\renewcommand{\arraystretch}{1.3}
\begin{tabular}{ lcccc }
\hline\hline
Par. & $c_0$ & $c_1$ & $c_2$ & $c_3$\\ \hline
A & -1.131$\times 10^{-4}$ & 2.939$\times 10^{-6}$ & 1.135$\times 10^{-8}$ & 4.935$\times 10^{-9}$  \\
\hline\hline
\end{tabular}
\end{table}

\subsection{Closure Approximation Fits}
\label{sec:triumf_closure_fits}

Before fitting the TRIUMF data with the phase-space approximation introduced in Sec.~\ref{sec:plestid_approx}, we first analyze the data using the closure approximation.
This serves two purposes:
First, it allows for a baseline comparison of the results obtained using the published and refitted detector responses with those reported by TRIUMF.
Second, it provides a direct assessment of how well the closure and phase-space approximations describe the high-energy tail of the RMC photon spectrum.

For comparison with the original TRIUMF analysis, we perform a $\chi^2$ fit in which the closure endpoint, \kmax, and overall normalization are freely varied, with no background component included.
Both the choice of a $\chi^2$ minimization and the omission of background contributions ensure consistency with the original TRIUMF fitting procedure.
The resulting endpoint values are compared to the published results in Table~\ref{tab:triumf_closure_results}.
Notably, the values of \kmax obtained using either the restricted fitting range or the alternative response model are systematically closer to the TRIUMF values than those obtained with the nominal response model and fitting range.
The larger disagreement observed with the nominal response further suggests that the detector-response treatment used in the original TRIUMF analysis differs from the published detector response.

The uncertainties obtained here are systematically smaller than the published values.
Based on Fig.~5.1 of Ref.~\cite{RMC_1995_Bergbusch_MS_thesis}, the TRIUMF uncertainties appear to have been determined using $\Delta\chi_\nu^2 = 1$, where $\chi^2_\nu$ is the reduced chi-squared metric for $\nu$ degrees of freedom, rather than $\Delta\chi^2=1$. This inflates the estimated uncertainties by approximately $\sqrt{\nu}$.
The fitted closure endpoints also lie well below the corresponding no-knockout kinematic limits listed in Table~\ref{tab:knockout_endpoint_energies}.
The resulting truncation of the spectrum illustrates the potential for the closure approximation to significantly underestimate the high-energy tail of the RMC photon spectrum.

We additionally fit the closure approximation using the same likelihood-based framework employed for the phase-space model in the following section.
Several datasets contain high-energy events that are unlikely to arise from RMC and are instead attributed to RPC and cosmic-ray contamination.
This motivates the inclusion of a floating flat background component in the fits, which is assumed to follow the same detector acceptance curve as the RMC spectrum.
The theoretical RMC spectrum is convolved with the detector response, and the closure endpoint, closure normalization, and background normalization are simultaneously fitted without prior constraints.
The best-fit parameters are obtained using a maximum-likelihood fit in the ROOT~\cite{Brun:1997pa} RooFit package~\cite{Verkerke:2003ir}.

\begin{figure}
\centering
    \includegraphics[width=1.00\linewidth, trim = 0 0 0 0, clip]{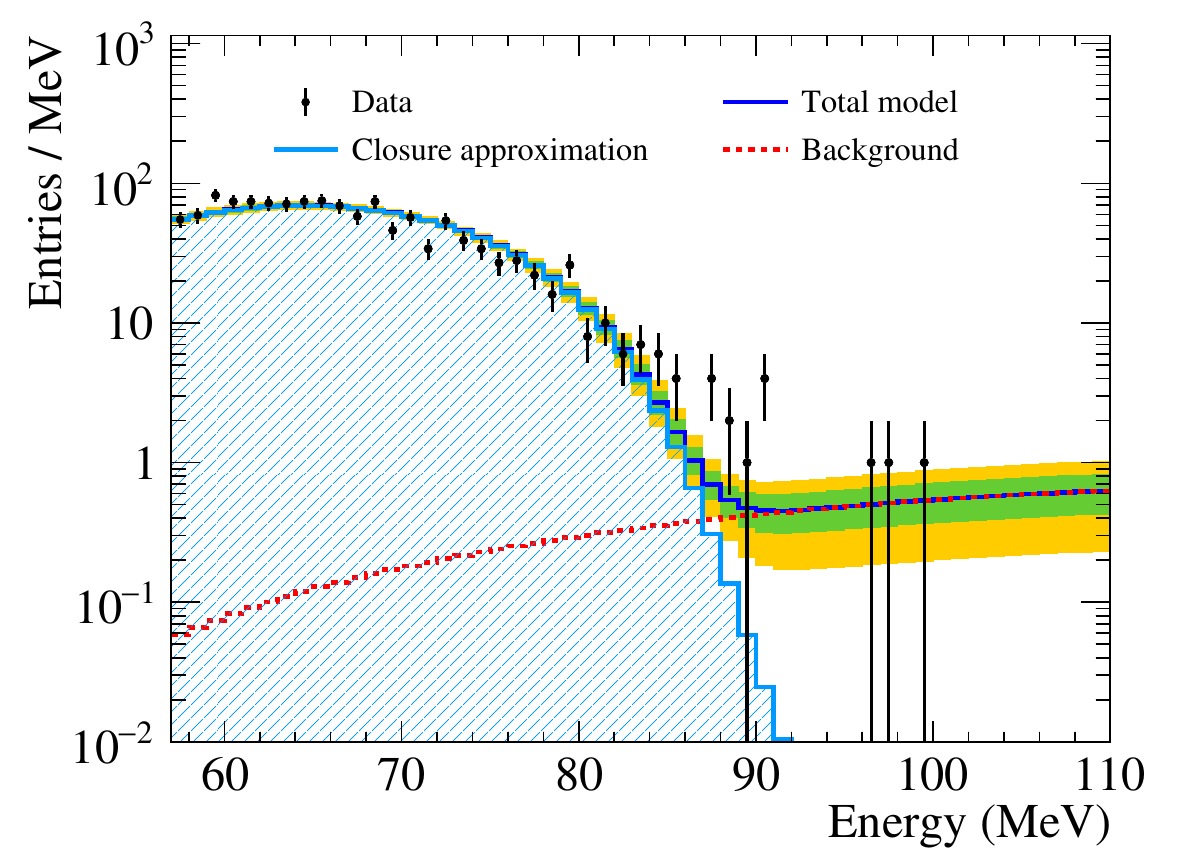}
    \caption{Closure approximation fit to the 1992 aluminum RMC data from Ref.~\cite{RMC_1992_PhysRevC.46.1094}, including a floating flat background component.
    }
    \label{fig:triumf_closure_log}
\end{figure}

The resulting fitted spectrum with the nominal detector response is shown in Fig.~\ref{fig:triumf_closure_log}.
Although the background component accounts for a few events above the fitted closure endpoint, the model fails to fully describe the observed high-energy tail.
This limitation directly motivates the phase-space fits presented in Sec.~\ref{sec:analysis_triumf_phasespace}.

\begin{table}[t]
  \caption{Closure-approximation endpoint values reported in Ref.~\cite{RMC_1992_PhysRevC.46.1094} ($k_{\max}^{\mathrm{pub}}$) compared with those obtained using the published detector response over the full energy range (\kmax), the refitted response (\kmaxstar), and the published response with the fit restricted to $E>65~\mathrm{MeV}$ (\kmaxstarstar).}
  \label{tab:triumf_closure_results}
  \centering
  \renewcommand{\arraystretch}{1.2}
  \setlength{\tabcolsep}{4.5pt}
  \begin{tabular}{ccccc}
    \hline\hline
    Nucleus & $k_{\max}^{\mathrm{pub}}$ & \kmax & \kmaxstar & \kmaxstarstar \\
    \hline
    Al & $90.0 \pm 2.0$ & $88.6 \pm 0.5$ & $89.5 \pm 0.5$ & $89.7 \pm 0.6$ \\
    Si & $92.0 \pm 2.0$ & $89.7 \pm 0.4$ & $90.6 \pm 0.4$ & $91.6 \pm 0.5$ \\
    Ca & $93.0 \pm 2.0$ & $91.4 \pm 0.4$ & $92.2 \pm 0.4$ & $92.2 \pm 0.5$ \\
    Mo & $90.0 \pm 2.0$ & $88.5 \pm 0.5$ & $89.4 \pm 0.4$ & $90.0 \pm 0.5$ \\
    Sn & $87.0 \pm 2.0$ & $85.7 \pm 0.5$ & $86.7 \pm 0.5$ & $89.1 \pm 0.7$ \\
    Pb & $84.0 \pm 3.0$ & $82.5 \pm 0.5$ & $83.4 \pm 0.5$ & $84.4 \pm 0.7$ \\
    \hline\hline
  \end{tabular}
\end{table}

\subsection{Phase-Space Approximation Fits}
\label{sec:analysis_triumf_phasespace}
The phase-space approximation is fitted to the data using a finite set of nuclear final states that are freely floated.
The kinematic endpoint energy for a nuclear final state with $a$ ejected neutrons and $b$ ejected protons is given by the effectively two-body decay formula 
\begin{equation}
    \kmax = \frac{M_i^2 - M_f^2}{2M_i}~\,,
\end{equation}
where $M_i = m_\mu + M_{N_i} - B_\mu$ is the initial system mass of the muon and the nucleus ($N_i$), $M_f = M_{N_f} + a\,m_n + b\,m_p$ is the mass of the recoiling nucleus ($N_f$) plus the mass of the ejected nucleons, and $B_\mu$ is the muon $1\mathrm{s}$ binding energy, which we take from Ref.~\cite{bindingEnergies}.
The neutrino masses are set to zero here.
The nuclear masses are evaluated using the atomic masses reported in Ref.~\cite{Wang_2017} with the contributions from atomic electrons removed.

We first consider zero-, one-, and two-nucleon-knockout final states.
Across all targets, only the zero-knockout and higher-endpoint one-nucleon-knockout channels have non-negligible contributions within the reconstructed energy range.
We therefore retain only these two channels in the final fits.
Table~\ref{tab:knockout_endpoint_energies} shows the endpoints for the different nuclear targets, assuming the most abundant isotope for each target.
In the cases of Al, Si, and Ca, the natural abundances are highly concentrated in a single isotope, while for Mo, Sn, and Pb there are non-negligible contributions from subleading isotopes.

The fit results are shown in Table \ref{tab:triumf_plestid_results}, where \br denotes each final state's fractional contribution to the RMC spectrum above $57\,\text{MeV}$.
The phase-space model yields lower $\chi^2_\nu$ values than the closure approximation for every dataset, where each fit has an identical number of degrees of freedom.
This modeling can be further improved by including contributions from additional isotopes for targets with non-negligible isotope mixtures, as such contributions alter the predicted photon spectrum.
This limitation does not apply to aluminum, the planned stopping target for COMET and Mu2e, which is monoisotopic.
Combining these fractions with the partial branching ratios $R_\gamma$ for photons above $57\,\text{MeV}$ measured in Ref.~\cite{RMC_1992_PhysRevC.46.1094} gives the per-channel partial branching ratios $R_{\gamma,i} = \br_i R_\gamma$ listed in Table~\ref{tab:triumf_partial_rates}.

\begin{table}
  \caption{Kinematic photon energy endpoints in MeV for each relevant nucleus by nuclear final state.}
  \label{tab:knockout_endpoint_energies}
  \centering
  \begin{tabular}{cccc}
  \hline\hline
    Nucleus & No knockout & Neutron knockout & Proton knockout \\
    \hline
    Al & $101.87$ & $95.45$ & $86.91$ \\
    Si & $99.78$  & $92.08$ & $90.26$  \\
    Ca & $102.63$ & $94.85$ & $95.07$  \\
    Ti & $99.80$  & $91.58$ & $90.37$  \\
    Mo & $96.58$  & $90.60$ & $88.72$  \\
    Sn & $94.51$  & $88.42$ & $85.48$  \\
    Au & $94.32$  & $88.48$ & $86.05$  \\
    Pb & $89.50$  & $85.72$ & $81.96$  \\
  \hline\hline
  \end{tabular}
\end{table}

\begin{figure}
\centering
    \includegraphics[width=0.99\linewidth, trim = 0 0 0 0, clip]{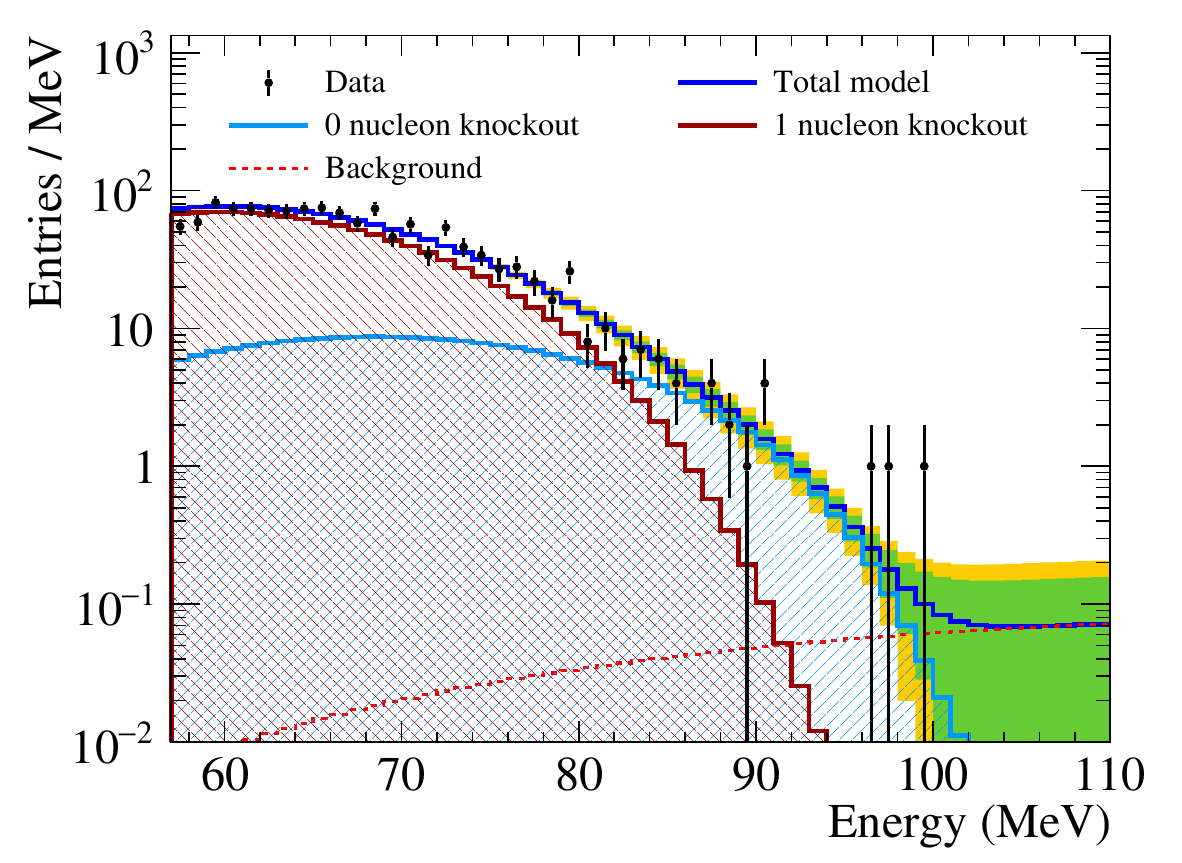}
    \caption{
    Phase-space approximation fit result to the TRIUMF aluminum data published in Ref.~\cite{RMC_1992_PhysRevC.46.1094}. The green (yellow) band represents the $1\sigma$ ($2\sigma$) fit statistical uncertainty bands.
    }
    \label{fig:triumf_plestid_fit_al}
\end{figure}

\begin{table}
  \caption{Phase-space fit results to the data published in Ref.~\cite{RMC_1992_PhysRevC.46.1094} by nuclear target with symmetric statistical uncertainties and asymmetric systematic uncertainties. \br values are given as percentages and $\chi^2_\nu$ is given for both the phase-space fits and the closure approximation (CA) fits for comparison.
  }
  \label{tab:triumf_plestid_results}
  \centering
  \renewcommand{\arraystretch}{1.3}
  \begin{tabular}{ccccc}
  \hline\hline
    Target &
    \br\!(0 knockout) &
    \br\!(1 knockout) &
    $\chi^2_{\nu}$ &
    $\chi^2_{\nu}$(CA) \\
    \hline
    Al &
    $9.9 \pm 1.9\,{}^{+4.4}_{-3.4}$ &
    $90.1 \pm 1.9\,{}^{+3.4}_{-4.4}$ &
    $1.27$ &
    1.61 \\
    Si &
    $27.3 \pm 2.2\,{}^{+7.5}_{-4.7}$ &
    $72.7 \pm 2.2\,{}^{+4.7}_{-7.5}$ &
    $1.70$ &
    3.48 \\
    Ca &
    $25.0 \pm 2.1\,{}^{+6.8}_{-8.3}$ &
    $75.0 \pm 2.1\,{}^{+8.3}_{-6.8}$ &
    $1.73$ &
    2.22 \\
    Mo &
    $31.8 \pm 3.1\,{}^{+9.9}_{-7.3}$ &
    $68.2 \pm 3.1\,{}^{+7.3}_{-9.9}$ &
    $1.13$ &
    1.59 \\
    Sn &
    $26.0 \pm 4.6\,{}^{+7.9}_{-0.5}$ &
    $74.0 \pm 4.6\,{}^{+0.5}_{-7.9}$ &
    $0.61$ &
    2.57 \\
    Pb &
    $27.4 \pm 4.1\,{}^{+11.3}_{-4.5}$ &
    $72.6 \pm 4.1\,{}^{+4.5}_{-11.3}$ &
    $0.66$ &
    1.23 \\
  \hline\hline
  \end{tabular}
\end{table}

\begin{table}
  \caption{Partial RMC branching ratios per ordinary muon capture for photons above $57\,\text{MeV}$, $R_{\gamma,i}=\br_i R_\gamma$, in units of $10^{-5}$, using $R_\gamma$ from Ref.~\cite{RMC_1992_PhysRevC.46.1094}. The first uncertainty combines the fit statistical uncertainty on $\br_i$ with the published uncertainty on $R_\gamma$; the second is the systematic uncertainty on $\br_i$.}
  \label{tab:triumf_partial_rates}
  \centering
  \renewcommand{\arraystretch}{1.3}
  \begin{tabular}{cccc}
  \hline\hline
    Target & $R_\gamma$ & $R_{\gamma,0}$ (0 knockout) & $R_{\gamma,1}$ (1 knockout) \\
    \hline
    Al & $1.43 \pm 0.13$ & $0.14 \pm 0.03\,{}^{+0.06}_{-0.05}$ & $1.29 \pm 0.12\,{}^{+0.05}_{-0.06}$ \\
    Si & $1.93 \pm 0.18$ & $0.53 \pm 0.06\,{}^{+0.14}_{-0.09}$ & $1.40 \pm 0.14\,{}^{+0.09}_{-0.14}$ \\
    Ca & $2.09 \pm 0.19$ & $0.52 \pm 0.06\,{}^{+0.14}_{-0.17}$ & $1.57 \pm 0.15\,{}^{+0.17}_{-0.14}$ \\
    Mo & $1.11 \pm 0.11$ & $0.35 \pm 0.05\,{}^{+0.11}_{-0.08}$ & $0.76 \pm 0.08\,{}^{+0.08}_{-0.11}$ \\
    Sn & $0.98 \pm 0.09$ & $0.25 \pm 0.05\,{}^{+0.08}_{-0.01}$ & $0.73 \pm 0.08\,{}^{+0.01}_{-0.08}$ \\
    Pb & $0.60 \pm 0.07$ & $0.16 \pm 0.03\,{}^{+0.07}_{-0.03}$ & $0.44 \pm 0.06\,{}^{+0.03}_{-0.07}$ \\
  \hline\hline
  \end{tabular}
\end{table}

\subsection{Bayesian Unfolding}

In addition to forward convolution fits, we employ unfolding methodologies to evaluate the RMC photon spectrum shape. Unfolding yields a model-independent estimate of the true spectrum, reducing reliance on a priori theoretical assumptions. Furthermore, this approach is highly computationally efficient; fitting arbitrary theoretical models to the unfolded data bypasses the need for expensive, repeated forward convolutions.

However, unfolding is not typically implemented due to difficulties associated with direct matrix inversion on the detector response that can produce highly unstable results. This in turn means that fluctuations in RMC data can lead to unphysical oscillations in the unfolded spectrum. To mitigate these instabilities, we employ deconvolution through the Richardson--Lucy algorithm~\cite{Schmitt_2017},
\begin{equation}
x_j^{(n+1)} = \frac{x_j^{(n)}}{\varepsilon_j} \sum_i R_{ij} \frac{d_i}{\sum_k R_{ik} x_k^{(n)}} \ ,
\end{equation}
where $x_j^{(n+1)}$ is the updated estimate of the true spectrum for the energy bin $j$ at the next iteration, $x_j^{(n)}$ is the current estimate of the true spectrum at iteration $n$, $d_i$ is the number of observed counts in reconstructed bin $i$, $R_{ij}$ is the detector response matrix element representing the probability of a true event in bin $j$ being reconstructed in bin $i$,
and $\varepsilon_j = \sum_i R_{ij}$ is the selection and detection efficiency for true bin $j$.

The Richardson--Lucy algorithm is particularly advantageous as its iterative nature suppresses noise amplification by gradually converging toward the true underlying distribution. It is also based on Bayesian principles, so it acts as a maximum likelihood estimator for Poisson statistics. Furthermore, its multiplicative structure naturally enforces a positivity constraint on the resulting counts, which is physically necessary for RMC studies. The detector response published in Ref.~\cite{RMC_1992_PhysRevC.46.1094} is used as the primary model for testing purposes.

However, unfolding artifacts can still emerge near the edges of the data spectrum, particularly in the low-energy region where the detector acceptance drops rapidly. To mitigate these edge effects, we evaluate two approaches: artificially extending the deconvolution region using simulated uniform counts on an expanded energy grid, and truncating the edge data post-unfolding. While both methods prove effective in toy studies, we ultimately choose to truncate the spectrum to also exclude the ill-conditioned region below $65\,\text{MeV}$. These tests demonstrate that removing this region eliminates the edge artifacts, resulting in significantly more stable and unbiased parameter extractions.

Following the deconvolution of the raw data, a rigorous evaluation of the statistical uncertainties is necessary. To estimate these uncertainties and bin-to-bin correlations introduced through unfolding, we use a Monte Carlo error propagation technique. A straightforward implementation generates the ensemble of toy datasets by Poisson-sampling directly from the measured data and re-unfolding each replicate. However, closure testing with a known injected truth reveals that this approach is self-referential in the near-singular, collinear region of the response matrix: the direction in which a noise realization shifts the unfolded point estimate also biases the covariance matrix subsequently used to fit that same realization, which reinforces rather than corrects the excursion. This produces a branching ratio that is biased low relative to the injected truth; in a sizable fraction of pseudo-experiments, it can collapse the fit outright to the non-negative boundary.

To correct this, the covariance is instead estimated iteratively. Following an initial fit, the Poisson ensemble is regenerated around the current best-fit model prediction rather than the raw data, and the spectrum is refitted against this updated covariance. This process is repeated until the extracted parameters stabilize, typically after four to six repetitions. At each repetition, a large ensemble of toy datasets ($N = 10{,}000$) is generated by sampling from independent Poisson distributions centered on the current prediction and propagating each fluctuated dataset through the standard unfolding algorithm. We calculate the covariance matrix from the resulting ensemble of unfolded spectra, where the diagonal elements represent the variances of each energy bin, finally allowing us to quantify the error.

Once the spectrum has been unfolded and its uncertainties estimated, we use Bayesian analysis techniques with a Markov Chain Monte Carlo to extract parameter estimates. We use a Gaussian log-likelihood for computational efficiency. This is conducted in Julia using the Bayesian Analysis Toolkit (BAT.jl) package \cite{schulz2020batjljuliabasedtool}.
We fit the phase-space approximation presented in Sec.~\ref{sec:plestid_approx} to the unfolded data.
Because the primary advantage of analyzing unfolded data is that it does not require an assumed theoretical model, fitting specific physical models serves primarily as a cross-check. In this context, we initially float three distinct nucleon-knockout final states to evaluate their relative contributions. Consistent with Sec.~\ref{sec:analysis_triumf_phasespace}, we find that the lowest-energy endpoint has a negligible impact on the overall fit and is therefore omitted.

As we make no a priori assumption on the relative contributions of the distinct nucleon knockout final states, we use flat, uninformative priors in the parameter extraction.
The binned log-likelihood function uses a multivariate Gaussian to help account for the correlation introduced to unfolded data points due to the deconvolution process.

We validate the unfolding pipeline using pseudo-experiments designed to perform a closure test. These simulated datasets include a small, flat background component representing the cosmic-ray and RPC contamination expected in the experimental data. Incorporating this background naturally stabilizes low-count regions against severe statistical fluctuations, providing a more physically robust benchmark.

The number of Richardson--Lucy iterations is optimized using these closure tests by injecting a known branching ratio into the simulated data, propagating it through the full unfolding and fitting pipeline, and minimizing the resulting bias relative to the injected truth. Based on these pseudo-experiments, an iteration count of 10 is found to minimize parameter bias and is selected for the final analysis.

\begin{figure}[thb!]
    \centering
    \includegraphics[width=1.00\linewidth, trim = 0 0 0 0, clip]{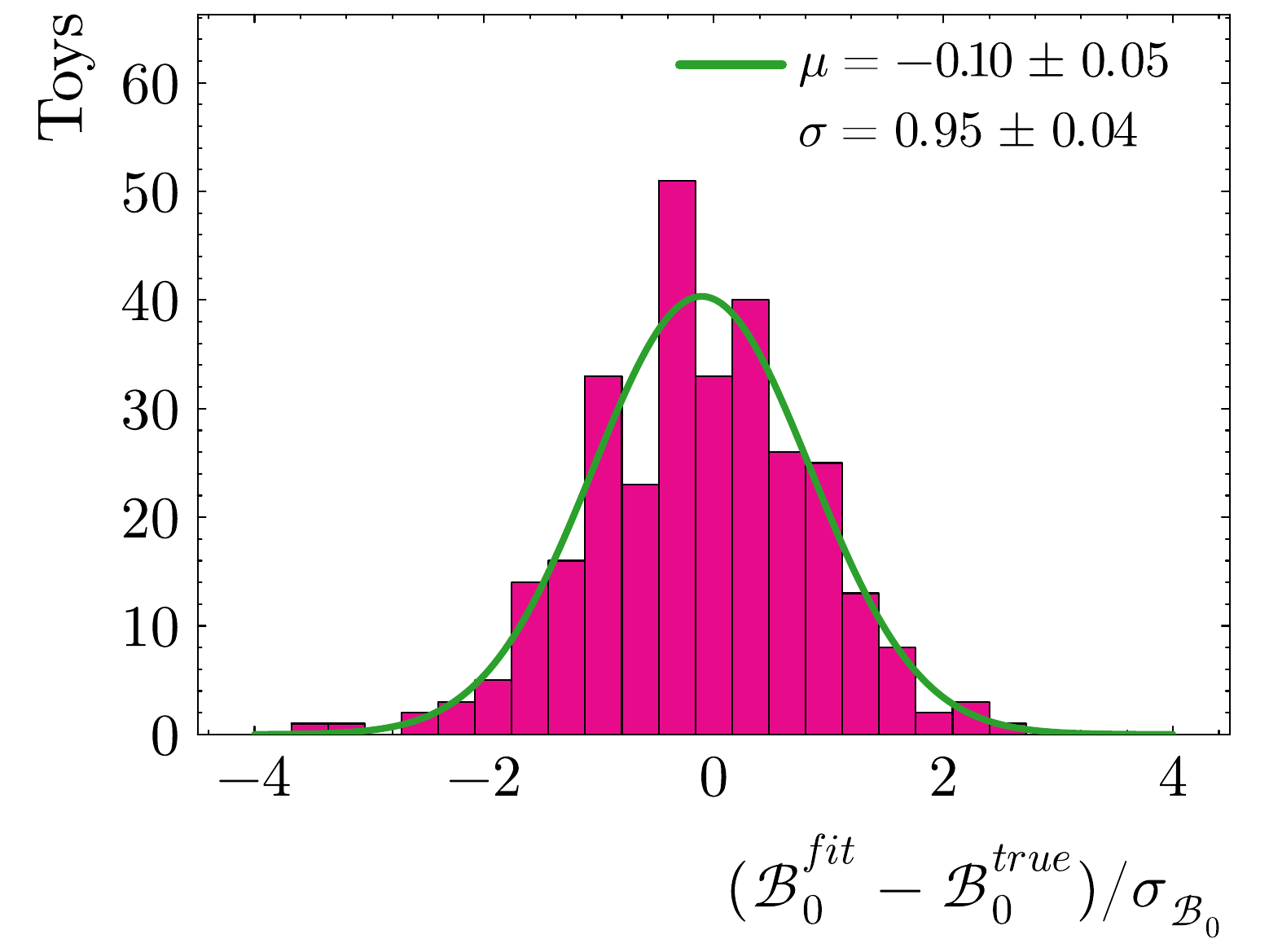}
    \caption{Pull distribution for the fitted $\br_0$ parameter, where the corresponding $\br_1$ pull distribution is a mirror of the $\br_0$ distribution. The green line shows a fitted Gaussian distribution, with the shape parameters given in the legend along with their statistical uncertainties.}
\label{fig:pull_test}
\end{figure}

The pull distribution for the fitted $\br_{0}$ value is shown in Fig.~\ref{fig:pull_test}. The distribution is consistent with a normal distribution, as expected. The pull mean is slightly offset from zero, but this residual bias is small in comparison to the statistical uncertainty.

This framework is then applied to the 1992 radiative muon capture data on $^{27}\text{Al}$ from Ref.~\cite{RMC_1992_PhysRevC.46.1094}.
We apply 10 unfolding iterations to the data, producing an unfolded spectrum from $50$ to $105\,\text{MeV}$, which is then truncated to include the range between $65$ and $100\,\text{MeV}$ for fitting purposes.
This region is chosen due to its stability in unfolding.

\begin{figure}[hbt!]
\centering
\includegraphics[width=1.00\linewidth, trim = 0 0 0 0, clip]{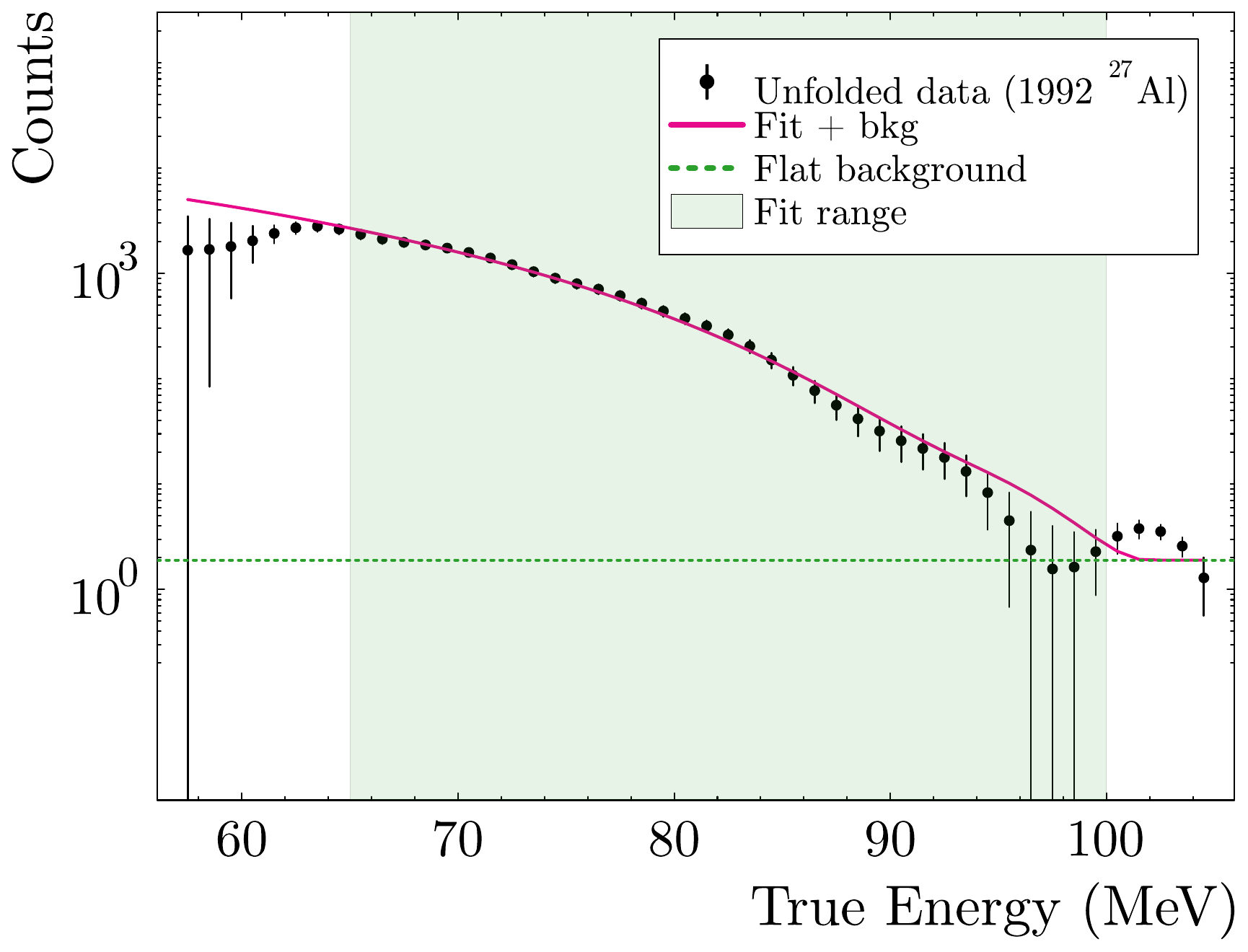}
\caption{The phase-space approximation model fitted to the unfolded aluminum data with a background component found to be $1.9$ counts/MeV, compared against the unfolded aluminum data. The shaded band marks the fit range, $65$--$100\,\text{MeV}$, drawn from the first to the last bin edge used in the fit.}
\label{fig:log_fit_1992}
\end{figure}

The effective branching ratios extracted from our fits for the radiative muon capture knockout channels are $\br_{0} = 8.44 ^{+1.66}_{-1.64}\%$ and $\br_{1} = 91.56 ^{+1.64}_{-1.66}\%$, where the uncertainties quoted correspond to the $68\%$ credible intervals obtained from the posterior distributions.
These results are consistent with the forward convolution fit results shown in Table~\ref{tab:triumf_plestid_results} within the statistical uncertainties.
The fitted model is shown against the unfolded data in Fig.~\ref{fig:log_fit_1992}.
The artifact in the high-energy tail is due to fluctuations in the observed data.

Ultimately, it is found that the integration of the Richardson--Lucy unfolding algorithm prior to Bayesian fitting provides a stable and computationally efficient framework for analyzing RMC spectra. With uncertainties estimated through Monte Carlo simulations and validated with pseudo-experiments, this method properly mitigates detector resolution effects and describes expected bin-to-bin correlations.
Applying this framework to the $^{27}\text{Al}$ dataset allows us to successfully extract the underlying kinematics, showing a spectrum heavily dominated by the single-neutron knockout channel. This demonstrates the ability of this method to produce reliable supplemental results for future RMC experiments.
It is also important to note that while the multivariate Gaussian log-likelihood approach attempts to address the fact that unfolding inherently introduces bin-to-bin correlations, which in turn can artificially correlate the extracted physical parameters, fully disentangling these effects remains a complex challenge that may require future investigations.

%% file: analysis_sindrum.tex
\section{Analysis of SINDRUM~II Data}
\label{sec:analysis_sindrum}

We digitize the historical SINDRUM~II muon-conversion search data from Refs.~\cite{1993_CLFV_SINDRUM_II_TITANIUM,1997_CLFV_SINDRUM_II_Kaulard_Thesis,2006_CLFV_SINDRUM_II_GOLD}. Because the original publications did not provide an analytic detector response function, we construct one using published simulation and data distributions.
Using the constructed response model, we fit models of the RMC photon spectrum to the observed high-energy positron distribution produced by asymmetric real and virtual photon conversions.
This analysis builds on the work in Ref.~\cite{RMC_2020_SINDRUMII_Excess_arXiv_2009.00214}, which analyzed the SINDRUM~II positron data in the context of \mumep conversion and first reported the inability of the closure approximation to describe a data-excess in the high-energy tail.

\subsection{SINDRUM~II Detector Response}

\begin{figure}
    \centering
    \includegraphics[width=0.99\linewidth, trim = 0 0 0 0, clip]{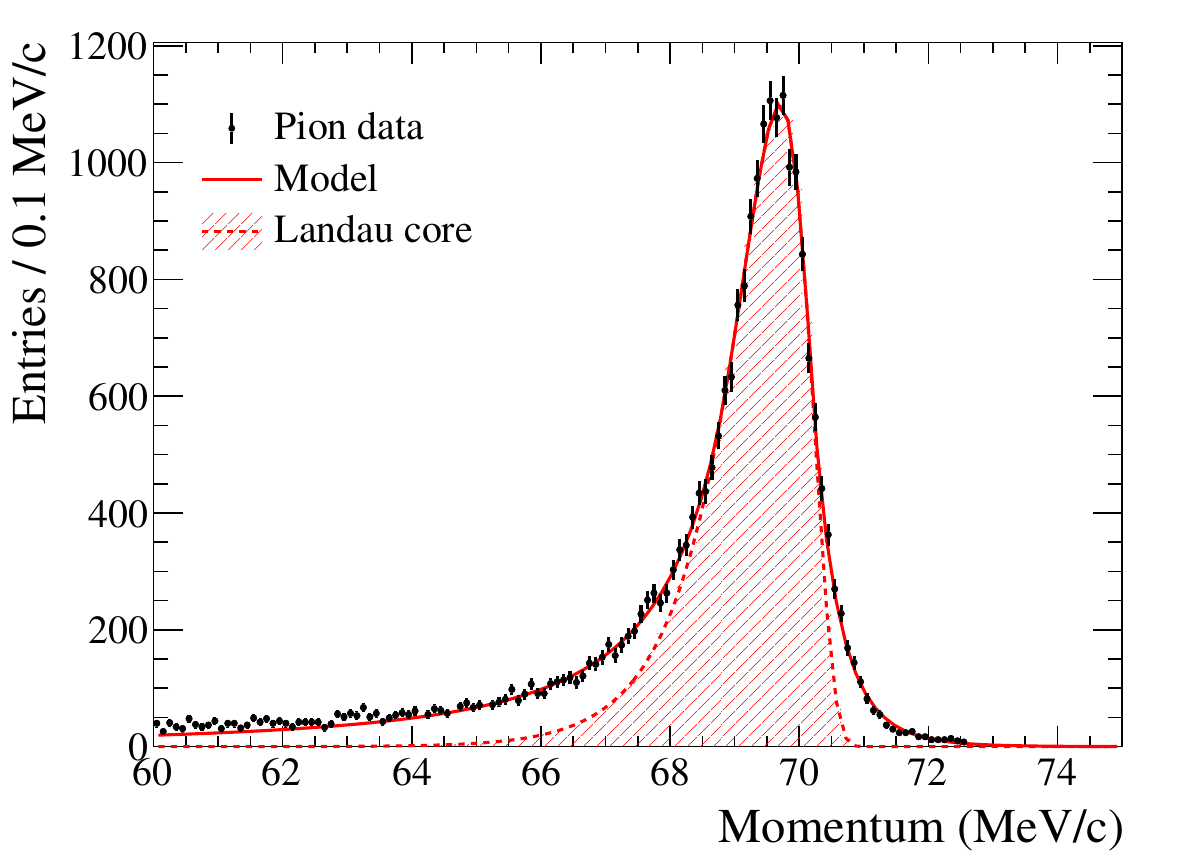}
    \caption{Pion calibration data from Ref.~\cite{SINDRUMII:1998mwd} fit with a double-sided Crystal Ball function~\cite{Skwarnicki:1986xj} with a Landau core.
    The dashed line with striped fill shows the core Landau distribution shape.
    }
    \label{fig:sindrum_pion_calib}
\end{figure}

\begin{figure}
    \centering
    \includegraphics[width=0.99\linewidth, trim = 0 0 0 0, clip]{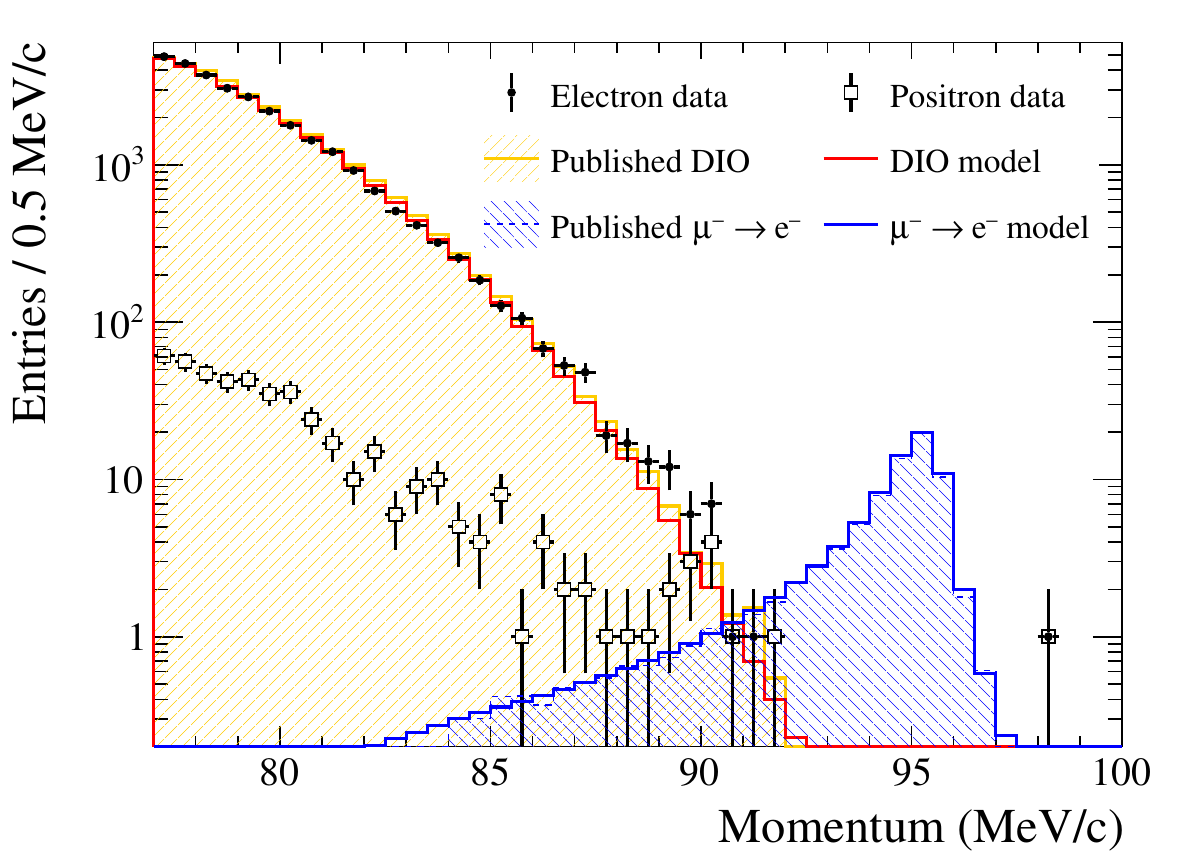}
    \caption{The DIO and \mumem models along with the published simulations overlaid with the measured electron and positron data from Ref.~\cite{2006_CLFV_SINDRUM_II_GOLD}.
    }
    \label{fig:sindrum_model}
\end{figure}

We construct a model of the detector resolution using the published SINDRUM~II $\pi^{+}\rightarrow e^{+}\nu_e$ calibration data.
We fit the digitized data with a double-sided Crystal Ball function~\cite{Skwarnicki:1986xj} that uses a Landau core instead of a Gaussian core, since the intrinsic resolution of the tracking is small compared with the energy losses at SINDRUM~II. 
The fit is performed in the $65\text{--}73\,\MeVc$ region, and the result is shown in Fig.~\ref{fig:sindrum_pion_calib}.
The fitted resolution model describes the calibration data well and is consistent with the published \mumem monoenergetic signal simulation, as shown in Fig.~\ref{fig:sindrum_model}.
In this work, we assume that the resolution is constant over the relevant momentum range and independent of charge.

\begin{figure}
    \centering
    \includegraphics[width=0.99\linewidth, trim = 0 0 0 0, clip]{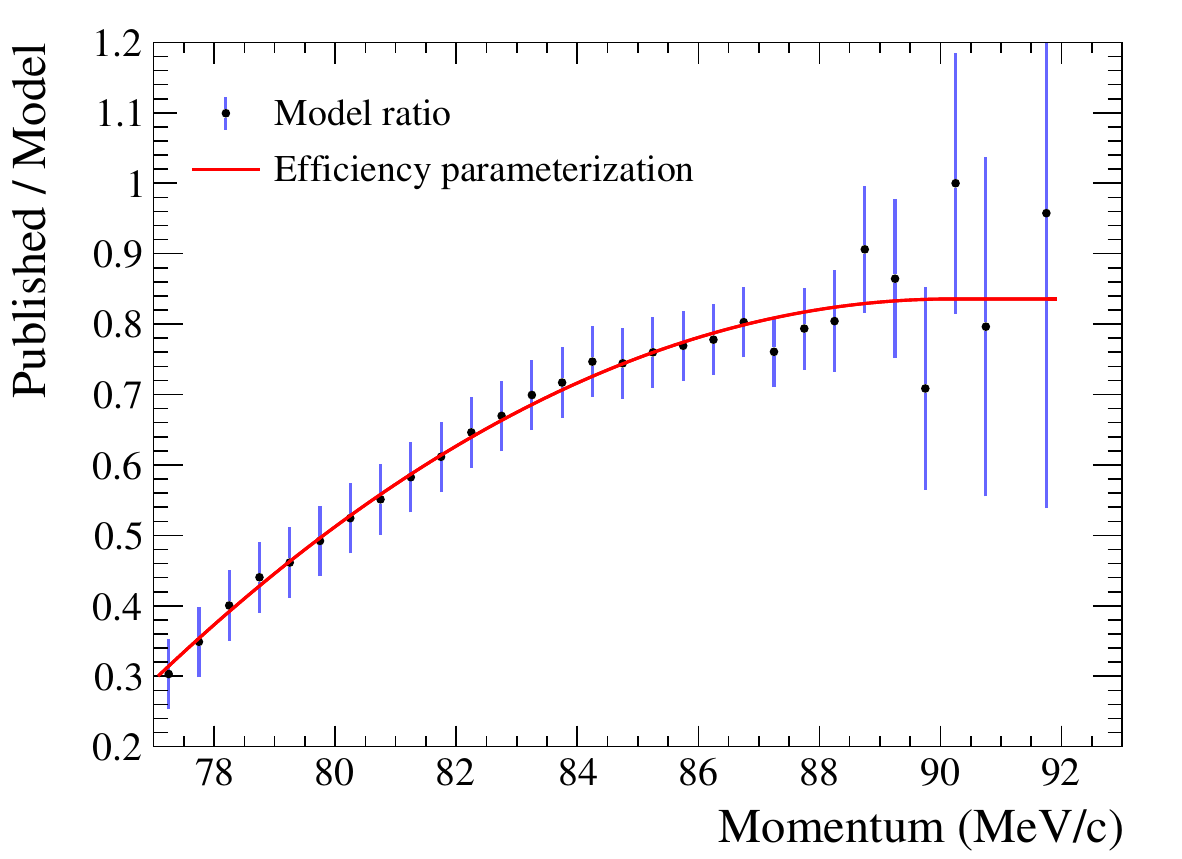}
    \caption{The plot shows the ratio of the published DIO model to the convolution of the theoretical DIO spectrum with the fitted detector resolution function, taken as the total efficiency function.
    The error bars are set to be the larger of 5\% and the Poisson bin uncertainty in the published model, assuming the distribution was created using a factor of 10 more simulated DIO events than it models.
    }
    \label{fig:sindrum_eff}
\end{figure}

We construct the detector acceptance function using the DIO simulation published in Ref.~\cite{2006_CLFV_SINDRUM_II_GOLD}.
To isolate the acceptance, we reproduce the theoretical DIO spectrum used in the original analysis so that any remaining difference after accounting for the detector resolution can be attributed to acceptance effects.
The gold DIO energy spectrum is taken as the lead DIO energy spectrum published by Watanabe et al.~\cite{Watanabe:1993}, shifted by the difference between the lead and gold endpoint energies ($0.54\,\text{MeV}$)~\cite{2006_CLFV_SINDRUM_II_GOLD}.
We interpolate the tabulated rates using the functional form adopted by SINDRUM~II in Ref.~\cite{2002_CLFV_SINDRUM_II_Annual_Report},
\begin{equation}
\frac{\dd \Gamma}{\dd E} \propto (E_{\text{max}} - E)^{5}e^{kE}\,,
\end{equation}
where $k = -0.1968\,\text{MeV}^{-1}$.
We convolve the resulting theoretical spectrum with the detector resolution model and fit the ratio of the published DIO simulation to the convolved spectrum to obtain the momentum-dependent acceptance function.
The resulting ratio and fitted acceptance curve are shown in Fig.~\ref{fig:sindrum_eff}.
We define the full detector response model as the product of the resolution and acceptance functions.
The resulting DIO model is shown overlaid with the published DIO simulation in Fig.~\ref{fig:sindrum_model}.

\subsection{RMC Modeling}

The SINDRUM~II spectrometer was optimized to reconstruct high-energy electron and positron tracks.
As a result, the RMC photon spectrum is accessible only indirectly via photon pair conversion and Compton scattering.
To reduce charged-particle energy loss, the target thicknesses and amount of detector material were kept small.
While this design improves tracking resolution, it also suppresses the rate of real photon conversion in material.
Consequently, virtual photon conversions contribute significantly to the observed RMC positron yield.

Using the relation derived in Ref.~\cite{PlestidAndHill}, we find that the positron spectra from real- and virtual-photon conversions have similar shapes at a given photon energy.
The difference in the positron spectra is small enough that the RMC positron distribution can be approximated solely from the real-photon conversion spectrum.
Additionally, we assume that the acceptance and resolution functions are the same as for target DIO electrons, neglecting subleading differences in the angular distributions of photon conversions in the target material and DIO.

\subsection{Closure Approximation Fits}
\label{sec:sindrum_closure}

\begin{figure}
    \centering
    \includegraphics[width=0.99\linewidth]{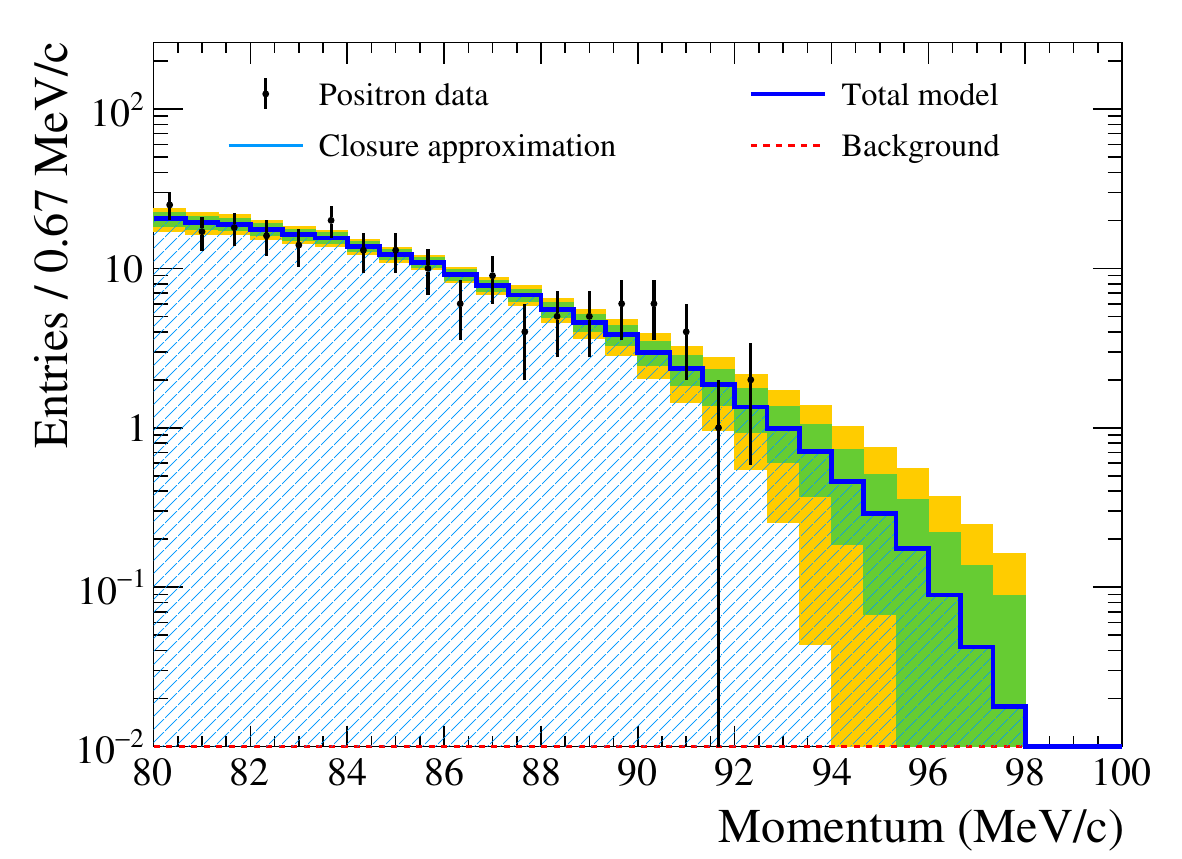}
    \caption{Closure approximation fit to the titanium SINDRUM~II positron dataset. The green (yellow) band represents the $1\sigma$ ($2\sigma$) fit statistical uncertainty bands.
    }
    \label{fig:sindrum_closure_ti}
\end{figure}

\begin{figure}
    \centering
    \includegraphics[width=0.99\linewidth]{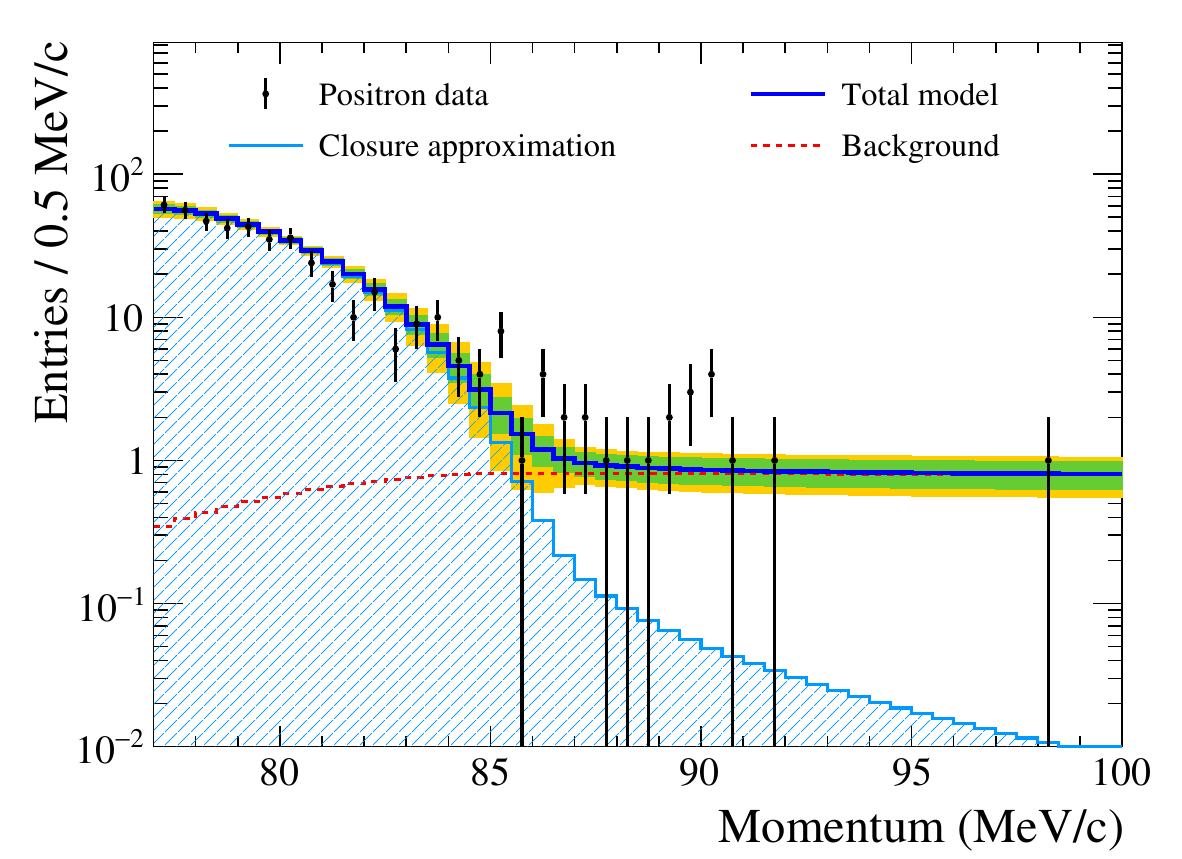}
    \caption{Closure approximation fit to the gold SINDRUM~II positron dataset. The green (yellow) band represents the $1\sigma$ ($2\sigma$) fit statistical uncertainty bands.
    }
    \label{fig:sindrum_closure_au}
\end{figure}

We fit the closure approximation to the SINDRUM~II titanium~\cite{SINDRUMII:1998mwd} and gold~\cite{2006_CLFV_SINDRUM_II_GOLD} positron datasets together with a freely floating flat background component to account for cosmic-ray and RPC contamination.
The resulting fits are shown in Figs.~\ref{fig:sindrum_closure_ti} and~\ref{fig:sindrum_closure_au}, while the fitted endpoints and fit qualities are summarized in Table~\ref{tab:sindrum_closure}.

For the titanium dataset, we obtain $\kmax = 99.2 \pm 1.2\,\text{MeV}$, well above the value of $\kmax = 89.2 \pm 2\,\text{MeV}$ reported in Ref.~\cite{RMC_1999_PhysRevC.59.2853}.
This inconsistency was noted in Ref.~\cite{SINDRUMII:1998mwd}, where a photon line was introduced at $93\,\text{MeV}$ to resolve the discrepancy.
This additional contribution was not physically motivated, but was instead chosen to resolve the tension with the data.
Because the signal region lay beyond this excess, a precise background model was not necessary for the published search.

For the gold dataset, the closure approximation fails to describe the high-energy tail, pulling the fitted background far above its expected level to accommodate events near and above the fitted endpoint.
Even with this inflated background contribution, the fit does not describe the high-energy excess previously noted in Ref.~\cite{RMC_2020_SINDRUMII_Excess_arXiv_2009.00214}.


\begin{table}[]
    \caption{Closure approximation fit results for the SINDRUM~II positron datasets.
    }
    \label{tab:sindrum_closure}
    \centering
    \renewcommand{\arraystretch}{1.1}
    \begin{tabular}{lcc}
    \hline\hline
        Target & \kmax & $\chi^2_\nu$ \\
        \hline
         Titanium & $99.2 \pm 1.2$ & 0.6 \\
         Gold     & $87.4 \pm 0.5$ & 1.7 \\ 
    \hline\hline
    \end{tabular}
\end{table}

\subsection{Phase-Space Approximation Fits}
\label{sec:sindrum_plestid}

\begin{figure}
    \centering
    \includegraphics[width=0.99\linewidth]{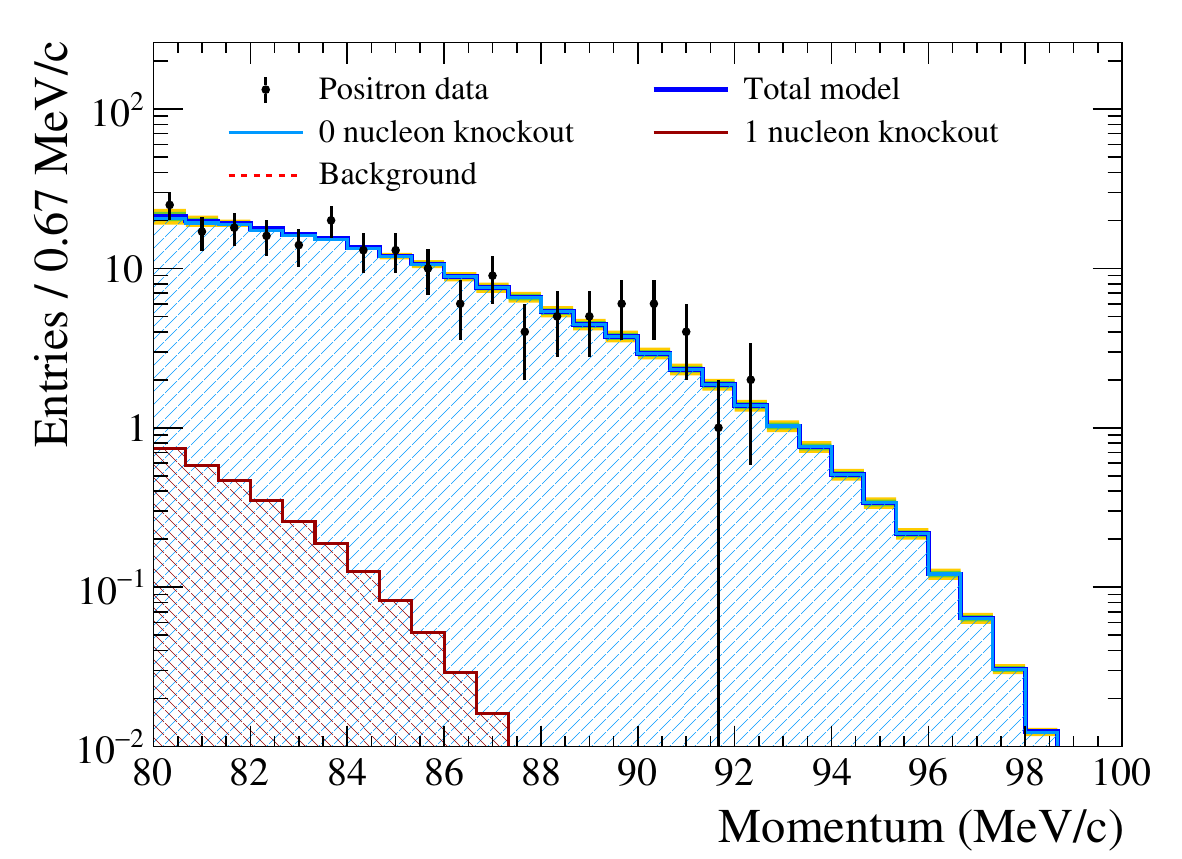}
    \caption{Phase-space approximation fit to the titanium SINDRUM~II positron dataset. The green (yellow) band represents the $1\sigma$ ($2\sigma$) fit statistical uncertainty bands.
    }
    \label{fig:sindrum_plestid_ti}
\end{figure}

\begin{figure}
    \centering
    \includegraphics[width=0.99\linewidth]{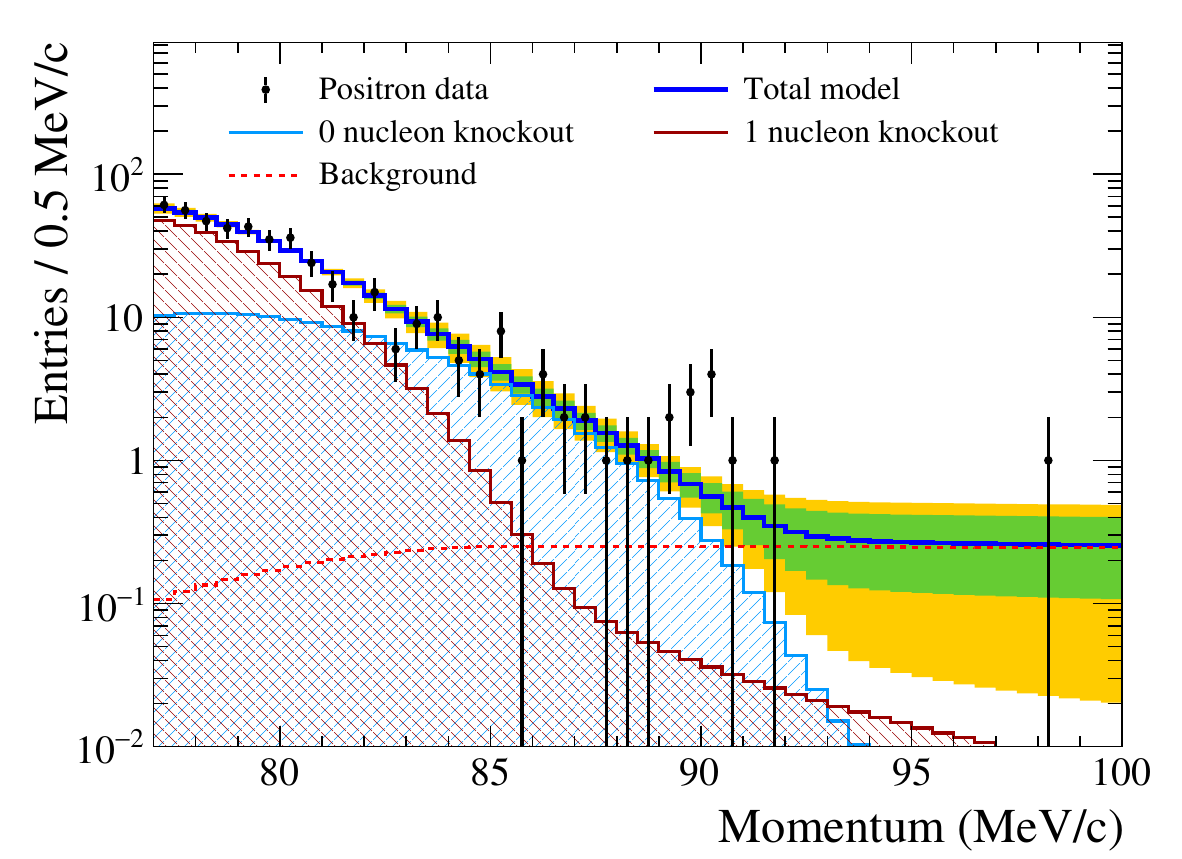}
    \caption{Phase-space approximation fit to the gold SINDRUM~II positron dataset. The green (yellow) band represents the $1\sigma$ ($2\sigma$) fit statistical uncertainty bands.
    }
    \label{fig:sindrum_plestid_au}
\end{figure}

We next fit the phase-space approximation to the SINDRUM~II positron data.
We model the spectrum using the two final states with the highest endpoints, together with a flat background component.
All three components are freely floated in the fits.
The results are shown in Figs.~\ref{fig:sindrum_plestid_ti} and~\ref{fig:sindrum_plestid_au}.

The phase-space approximation provides a better description of the data than the closure approximation, particularly for the gold target, where the excess above $85\,\text{MeV}$ is explained by the zero-nucleon-knockout contribution.
A notable peaking structure remains in the high-energy tail, though Ref.~\cite{RMC_2020_SINDRUMII_Excess_arXiv_2009.00214} found this feature to be inconsistent with a \mumep signal.
The titanium positron data are primarily modeled by the zero-nucleon-knockout final state, which has an endpoint energy value that is similar to the fitted closure approximation endpoint.
The phase-space approximation fit results are summarized in Table~\ref{tab:sindrum_plestid}.


\begin{table}[]
    \caption{Phase-space approximation fit results for the SINDRUM~II positron datasets.
    The branching fractions for the no nucleon and one nucleon knockout cases are given for E$_\gamma > 57$\,MeV to match the TRIUMF RMC dataset measurements, with asymmetric statistical uncertainties.
    }
    \label{tab:sindrum_plestid}
    \centering
    \renewcommand{\arraystretch}{1.3}
    \begin{tabular}{lccc}
    \hline\hline
        Target & $\mathcal{B}_0$ & $\mathcal{B}_1$ & $\chi^2_\nu$ \\
        \hline
         Titanium & $0.831^{+0.169}_{-0.444}$ & $0.169^{+0.444}_{-0.169}$ & 0.6 \\
         Gold     & $0.030^{+0.008}_{-0.008}$ & $0.969^{+0.008}_{-0.008}$ & 1.0 \\ 
    \hline\hline
    \end{tabular}
\end{table}

%% file: mu2e.tex
\section{Implications for Muon-Conversion Experiments}
\label{sec:rmc_implications}

 Our analyses of the TRIUMF RMC photon and SINDRUM~II positron spectra demonstrate that the closure approximation consistently fails to describe the observed high-energy tails.
 As the upcoming COMET and Mu2e experiments will collect $\mathcal{O}(10^4)$ times more stopped muons than the SINDRUM~II experiment, even a small RMC rate near the endpoint can produce a substantial background.
 The impact of the high-energy tail will depend on the target nucleus and the specific conversion channel. The aluminum RMC endpoint at $101.87\,\text{MeV}$ lies safely below the \mumem conversion energy of $104.97\,\text{MeV}$~\cite{endpoint_energy_paper}, but significantly exceeds the ground-state \mumep conversion threshold of $92.32\,\text{MeV}$~\cite{2022_CLFV_Lee}.
 While the high-energy RMC tail shape has a negligible effect on the projected \mumem sensitivity, the ultimate sensitivity of the \mumep channel is directly driven by the physical behavior of this high-energy tail.

We estimate the RMC positron background at COMET and Mu2e using the phase-space results for the TRIUMF aluminum data shown in Table~\ref{tab:triumf_partial_rates}.
As in SINDRUM~II, the experiments will be sensitive to both real and virtual photon conversions.
We use the relation given in Ref.~\cite{PlestidAndHill} to model the virtual conversion spectrum given the measured on-shell photon spectrum.
The photon spectrum fit results extracted in this work can therefore be used to predict the complete RMC background contributions to the \mumem and \mumep conversion searches at the COMET and Mu2e experiments.

Based on the COMET and Mu2e stopping-target designs, we estimate that a typical RMC photon traverses $\mathcal{O}(100\,\mu\text{m})$ of aluminum~\cite{bartoszek2015mu2etechnicaldesignreport,2020_COMET_Design_Report}.
Using the radiation length in aluminum, this corresponds to an external-conversion probability of $\mathcal{O}(10^{-3})$~\cite{PDG:2025_AluminumProps}.
Combining this with the internal-conversion contribution gives our total RMC positron background estimate.

\begin{figure}
    \centering
    \includegraphics[width=0.99\linewidth, trim = 0 0 0 0, clip]{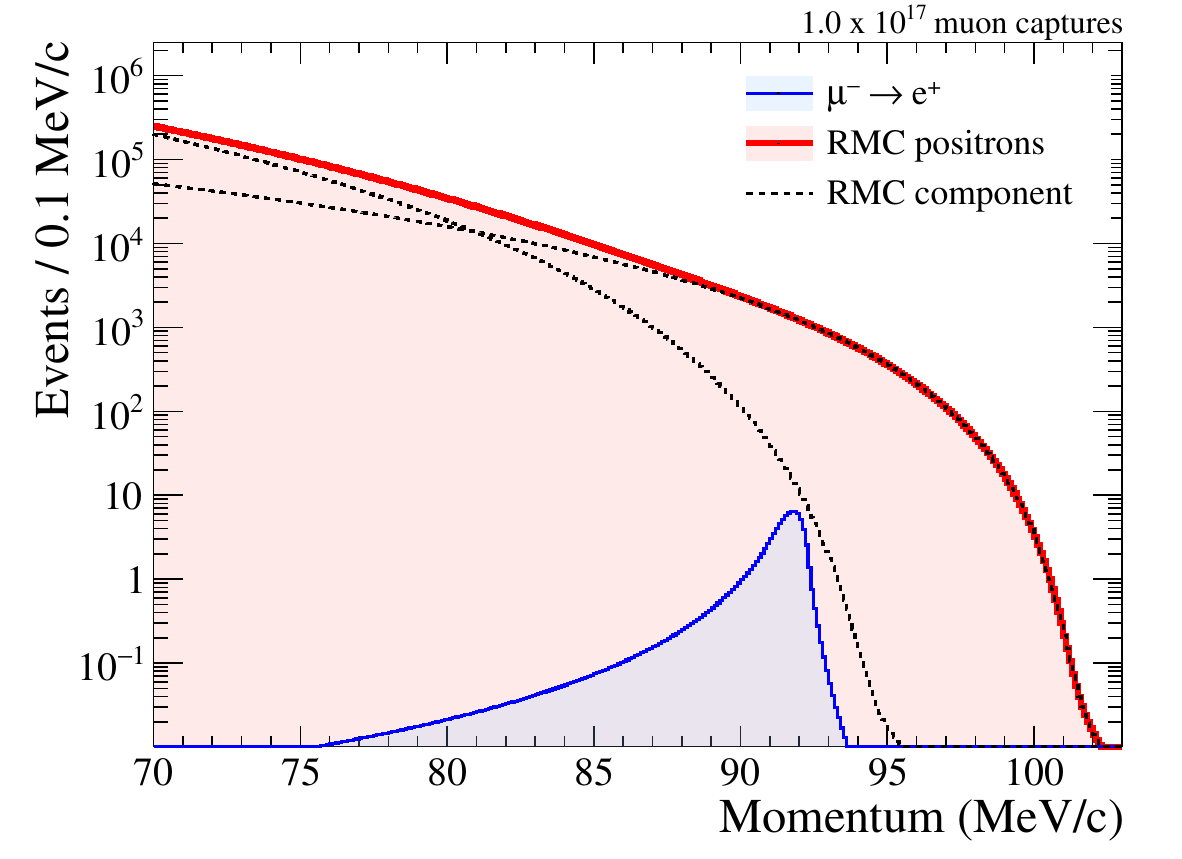}
    \caption{An example RMC positron background at a muon-conversion experiment for $10^{17}$ muon captures, using the phase-space approximation fit results. The dashed ``RMC component'' curves show the zero- and one-nucleon-knockout contributions (photon endpoints of $101.87$ and $95.45\,\text{MeV}$), with photon rates taken from Table~\ref{tab:triumf_partial_rates}, which sum to give the total spectrum (red line). A ground-state \mumep conversion signal with $\rmue = 10^{-14}$ is included for comparison.}
    \label{fig:mu2e_rmc_bkg_plestid}
\end{figure}

To estimate the reconstructed positron spectrum, we convolve the total RMC positron spectrum with a simplified detector response.
Guided by Ref.~\cite{Mu2e:2022Run1Sensitivity}, we assume an overall positron efficiency of $\mathcal{O}(10\%)$ and model the energy loss and momentum resolution using a Landau-core Crystal Ball function with a most probable energy loss of $\mathcal{O}(0.5\,\text{MeV})$.
Figure~\ref{fig:mu2e_rmc_bkg_plestid} shows the resulting spectrum for $10^{17}$ muon captures, with an example ground-state \mumep conversion signal overlaid for comparison.

The predicted background is $\mathcal{O}(10^4)$ events per \MeVc at the signal peak, dominated by the zero-nucleon-knockout component despite its smaller fractional contribution to the spectrum above $57\,\text{MeV}$.
The uncertainty on this prediction is therefore driven by the zero-knockout partial branching ratio, $R_{\gamma,0} = (0.14 \pm 0.03\,{}^{+0.06}_{-0.05})\times10^{-5}$, corresponding to an $\mathcal{O}(50\%)$ uncertainty on the background rate near the \mumep signal region.
Other backgrounds are expected to contribute $\lesssim 1$ count per~\MeVc, and are therefore negligible compared to RMC~\cite{bartoszek2015mu2etechnicaldesignreport,2020_COMET_Design_Report}.
Assuming the RMC background rate is constrained in situ, e.g., from the measured photon or positron spectra away from the signal region, we estimate a 90\% confidence-level sensitivity of $\rmue < \mathcal{O}(10^{-14}\text{--}10^{-13})$ for $10^{17}$ muon captures, where \rmue is the branching fraction of the CLFV conversion process relative to all muon captures on the target nucleus.
This would represent an improvement of one to two orders of magnitude over the previous SINDRUM~II search using a titanium target, which set an upper limit of $\rmue < 1.7\times10^{-12}$ at the 90\% confidence level~\cite{SINDRUMII:1998mwd}.

%% file: bib/bayesian_unfolding.bib
@article{Schmitt_2017,
   title={Data Unfolding Methods in High Energy Physics},
   volume={137},
   ISSN={2100-014X},
   url={http://dx.doi.org/10.1051/epjconf/201713711008},
   DOI={10.1051/epjconf/201713711008},
   journal={EPJ Web of Conferences},
   publisher={EDP Sciences},
   author={Schmitt, Stefan},
   editor={Foka, Y. and Brambilla, N. and Kovalenko, V.},
   year={2017},
   pages={11008} }

@misc{schulz2020batjljuliabasedtool,
      title={BAT.jl -- A Julia-based tool for Bayesian inference}, 
      author={Oliver Schulz and Frederik Beaujean and Allen Caldwell and Cornelius Grunwald and Vasyl Hafych and Kevin Kröninger and Salvatore La Cagnina and Lars Röhrig and Lolian Shtembari},
      year={2020},
      eprint={2008.03132},
      archivePrefix={arXiv},
      primaryClass={stat.CO},
      url={https://arxiv.org/abs/2008.03132}, 
}


%% file: bib/clfv_lnv_refs.bib
@article{Bernstein_2019,
   title={The Mu2e Experiment},
   volume={7},
   ISSN={2296-424X},
   url={http://dx.doi.org/10.3389/fphy.2019.00001},
   DOI={10.3389/fphy.2019.00001},
   journal={Frontiers in Physics},
   publisher={Frontiers Media SA},
   author={Bernstein, Robert H.},
   year={2019},
   month=jan 
}

@article{2020_COMET_Design_Report,
   title={COMET Phase-I technical design report},
   volume={2020},
   ISSN={2050-3911},
   url={http://dx.doi.org/10.1093/ptep/ptz125},
   DOI={10.1093/ptep/ptz125},
   number={3},
   journal={Progress of Theoretical and Experimental Physics},
   publisher={Oxford University Press (OUP)},
   author={Abramishvili, R and Adamov, G and Akhmetshin, R R and Allin, A and Ang{\'e}lique, J C and Anishchik, V and Aoki, M and Aznabayev, D and Bagaturia, I and Ban, G and Ban, Y and Bauer, D and Baygarashev, D and Bondar, A and others},
   year={2020},
   month=mar }

@misc{bartoszek2015mu2etechnicaldesignreport,
      title={Mu2e Technical Design Report}, 
      author={L. Bartoszek and E. Barnes and J. P. Miller and J. Mott and A. Palladino and J. Quirk and B. L. Roberts and J. Crnkovic and V. Polychronakos and V. Tishchenko and P. Yamin and C. -h. Cheng and others},
      year={2015},
      eprint={1501.05241},
      archivePrefix={arXiv},
      primaryClass={physics.ins-det},
      url={https://arxiv.org/abs/1501.05241}, 
}

@article{2006_CLFV_SINDRUM_II_GOLD,
author="Bertl, W.
        and Engfer, R.
        and Hermes, E. A.
        and Kurz, G.
        and Kozlowski, T.
        and Kuth, J.
        and Otter, G.
        and Rosenbaum, F.
        and Ryskulov, N. M.
        and van der Schaaf, A.
        and Wintz, P.
        and Zychor, I.
        and Collaboration, The SINDRUM II",
title    = "A search for $\mu$-e conversion in muonic gold",
journal  = "The European Physical Journal C - Particles and Fields",
year     = "2006",
month    = "Aug",
day      = "01",
volume   = "47",
number   = "2",
pages    = "337--346",
issn     = "1434-6052",
doi      = "10.1140/epjc/s2006-02582-x",
url      = "https://doi.org/10.1140/epjc/s2006-02582-x"
}

@article{1993_CLFV_SINDRUM_II_TITANIUM,
author="Dohmen, C.
        and Groth, K.
        and Heer, B.
        and Honecker, W.
        and Otter, G.
        and Steinrucken, B.
        and Wintz, P.
        and Djordjadze, V.
        and Hofmann, J.
        and Kolzowski, T.
        and Playfer, S.
        and Bertl, W.
        and Egger, J.
        and Herold, W.
        and Krause, B.
        and Walter, H.K.
        and Engfer, R.
        and Findeisen, Ch.
        and Grossmann-Handschin, M.
        and Hermes, E.A.
        and Muheim, F.
        and Niebuhr, C.B.
        and Pruys, H.S.
        and Ricken, L.
        and Vermeulen, D.
        and van der Shaaf, A.
        and Collaboration, The SINDRUM II",
title    = "Test of lepton-flavour conservation in $\mu\rightarrow e$ conversion on titanium",
journal  = "Physics Letters B",
year     = "1993",
month    = "",
day      = "",
volume   = "317",
number   = "4",
pages    = "631--636",
issn     = "0370-2693",
doi      = "10.1016/0370-2693(93)91383-X",
url      = "https://doi.org/10.1140/0370-2693(93)91383-X"
}

@article{endpoint_energy_paper,
  title = {Muon decay in orbit: Spectrum of high-energy electrons},
  author = {Czarnecki, Andrzej and Garcia i Tormo, Xavier and Marciano, William J.},
  journal = {Phys. Rev. D},
  volume = {84},
  issue = {1},
  pages = {013006},
  numpages = {8},
  year = {2011},
  month = {Jul},
  publisher = {American Physical Society},
  doi = {10.1103/PhysRevD.84.013006},
  url = {https://link.aps.org/doi/10.1103/PhysRevD.84.013006}
}

@article{SINDRUMII:1998mwd,
    author = "Kaulard, J. and others",
    collaboration = "SINDRUM II",
    title = "{Improved limit on the branching ratio of mu- --{\ensuremath{>}} e+ conversion on titanium}",
    doi = "10.1016/S0370-2693(97)01423-8",
    journal = "Phys. Lett. B",
    volume = "422",
    pages = "334--338",
    year = "1998"
}

@article{2022_CLFV_Lee,
AUTHOR = {Lee, MyeongJae and MacKenzie, Michael},
TITLE = {Muon to Positron Conversion},
JOURNAL = {Universe},
VOLUME = {8},
YEAR = {2022},
NUMBER = {4},
ARTICLE-NUMBER = {227},
URL = {https://www.mdpi.com/2218-1997/8/4/227},
ISSN = {2218-1997},
DOI = {10.3390/universe8040227}
}


%% file: bib/radiative_muon_capture.bib
@misc{RMC_2020_SINDRUMII_Excess_arXiv_2009.00214,
      title={Search for $\mu^- \rightarrow e^+$ conversion: what can be learned from the SINDRUM-II positron data on a gold target}, 
      author={MacKenzie, Michael and Murat, Pavel},
      year={2020},
      eprint={2009.00214},
      archivePrefix={arXiv},
      primaryClass={hep-ph}
}

@article{RMC_1999_PhysRevC.59.2853,
  title = {Radiative muon capture on {O}, {A}l, {S}i, {T}i, {Z}r, and {A}g},
  author = {Bergbusch, P. C. and Armstrong, D. S. and Blecher, M. and Chen, C. Q.
  and Doyle, B. C. and Gorringe, T. P. and Gumplinger, P. and Hasinoff, M. D.
  and Jonkmans, G. and Macdonald, J. A. and Poutissou, J-M. and Poutissou, R.
  and Sigler, C. N. and Wright, D. H.},
  journal = {Phys. Rev. C},
  volume = {59},
  issue = {5},
  pages = {2853--2864},
  numpages = {0},
  year = {1999},
  month = {May},
  publisher = {American Physical Society},
  doi = {10.1103/PhysRevC.59.2853},
  url = {https://link.aps.org/doi/10.1103/PhysRevC.59.2853}
}

@article{RMC_1998_PhysRevC.58.1767,
  title = {Isotope dependence of radiative muon capture on the ${}^{58,60,62}${N}i isotopes},
  author = {Gorringe, T. P. and Armstrong, D. S. and Chen, C. Q. and Christy, E. and Doyle, B. C.
  and Gumplinger, P. and Fearing, H. W. and Hasinoff, M. D. and Kovash, M. A. and Wright, D. H.},
  journal = {Phys. Rev. C},
  volume = {58},
  issue = {3},
  pages = {1767--1776},
  numpages = {0},
  year = {1998},
  month = {Sep},
  publisher = {American Physical Society},
  doi = {10.1103/PhysRevC.58.1767},
  url = {https://link.aps.org/doi/10.1103/PhysRevC.58.1767}
}

@mastersthesis{RMC_1995_Bergbusch_MS_thesis,
  author = {Bergbusch, Paul C.},
  school = {University of British Columbia},
  year   = {1995},
  url    = {https://open.library.ubc.ca/media/download/pdf/831/1.0085569/2}
}

@article{RMC_1992_WRIGHT_1992_249,
title = "The {TRIUMF} radiative muon capture facility",
journal = "Nuclear Instruments and Methods in Physics Research Section A: Accelerators, Spectrometers, Detectors and Associated Equipment",
volume = "320",
number = "1",
pages = "249 - 262",
year = "1992",
issn = "0168-9002",
doi = "10.1016/0168-9002(92)90783-Z",
url = "http://www.sciencedirect.com/science/article/pii/016890029290783Z",
author = "D.H. Wright and S. Ahmad and D.S. Armstrong and G. Azuelos and 
W. Bertl and M. Blecher and C.Q. Chen and T.P. Gorringe and M.D. Hasinoff 
and R.S. Henderson and J.A. Macdonald and J.M. Poutissou and R. Poutissou 
and B.C. Robertson and D.G. Sample and A. Serna-Angel and G. Taylor and N.S. Zhang",
}

@article{RMC_1992_PhysRevC.46.1094,
  title = {Radiative muon capture on {A}l, {S}i, {C}a, {M}o, {S}n, and {P}b},
  author = {Armstrong, D. S. and Serna-Angel, A. and Ahmad, S. and
                  Azuelos, G. and Bertl, W. and Blecher, M. and Chen,
                  C. Q. and Depommier, P. and von Egidy, T. and
                  Gorringe, T. P. and Hasinoff, M. D. and Henderson,
                  R. S. and Larabee, A. J. and Macdonald, J. A. and
                  McDonald, S. C. and Poutissou, J-M. and Poutissou,
                  R. and Robertson, B. C. and Sample, D. G. and
                  Taylor, G. N. and Wright, D. H. and Zhang, N. S.},
  journal = {Phys. Rev. C},
  volume = {46},
  issue = {3},
  pages = {1094--1107},
  numpages = {0},
  year = {1992},
  month = {Sep},
  publisher = {American Physical Society},
  doi = {10.1103/PhysRevC.46.1094},
  url = {https://link.aps.org/doi/10.1103/PhysRevC.46.1094}
}

@article{RMC_1991_PhysRevC.43.1425,
  title = {Radiative muon capture on carbon, oxygen, and calcium},
  author = {Armstrong, D. S. and Ahmad, S. and Burnham, R. A. and
                  Gorringe, T. P. and Hasinoff, M. D. and Larabee,
                  A. J. and Waltham, C. E. and Azuelos, G. and
                  Macdonald, J. A. and Numao, T. and Poutissou,
                  J-M. and Blecher, M. and Wright, D. H. and Clifford,
                  E. T. H. and Summhammer, J. and Depommier, P. and
                  Poutissou, R. and Mes, H. and Robertson, B. C.},
  journal = {Phys. Rev. C},
  volume = {43},
  issue = {3},
  pages = {1425--1447},
  numpages = {0},
  year = {1991},
  month = {Mar},
  publisher = {American Physical Society},
  doi = {10.1103/PhysRevC.43.1425},
  url = {https://link.aps.org/doi/10.1103/PhysRevC.43.1425}
}

@article{Christillin:1979fa,
      author         = "Christillin, P. and Rosa-Clot, M. and Servadio, S.",
      title          = "{Radiative Muon Capture in Medium Heavy Nuclei}",
      journal        = "Nucl. Phys.",
      volume         = "A345",
      year           = "1980",
      pages          = "331-366",
      doi            = "10.1016/0375-9474(80)90344-9",
      reportNumber   = "CERN-TH-2697",
      SLACcitation   = "%%CITATION = NUPHA,A345,331;%%"
}

@article{Christillin:1979pz,
      author         = "Christillin, P. and Rosa-Clot, M. and Servadio, S.",
      title          = "{Muon Wave Function and Coulomb Propagator in Radiative Muon Capture in Nuclei}",
      journal        = "Nucl. Phys.",
      volume         = "A345",
      year           = "1980",
      pages          = "317-330",
      doi            = "10.1016/0375-9474(80)90343-7",
      reportNumber   = "CERN-TH-2659",
      SLACcitation   = "%%CITATION = NUPHA,A345,317;%%"
}

@article{Christillin:1980nh,
    author = "Christillin, P.",
    title = "{Radiative Muon Capture for $N = Z$ Nuclei}",
    reportNumber = "CERN-TH-2862",
    doi = "10.1016/0375-9474(81)90501-7",
    journal = "Nucl. Phys. A",
    volume = "362",
    pages = "391--404",
    year = "1981"
}

@article{Primakoff:1959fs,
      author         = "Primakoff, H.",
      title          = "{Theory of muon capture}",
      journal        = "Rev. Mod. Phys.",
      volume         = "31",
      year           = "1959",
      pages          = "802-822",
      doi            = "10.1103/RevModPhys.31.802",
      SLACcitation   = "%%CITATION = RMPHA,31,802;%%"
}

@article{Fujii:1959aa,
author = {Fujii, A and Primakoff, H},
doi = {10.1007/BF02745906},
issn = {1827-6121},
journal = {Il Nuovo Cimento (1955-1965)},
number = {4},
pages = {327--355},
title = {{Muon capture in certain light nuclei}},
url = {https://doi.org/10.1007/BF02745906},
volume = {12},
year = {1959}
}

@article{Rood:1965aa,
author = {Rood, H P C and Tolhoek, H A},
doi = {https://doi.org/10.1016/0029-5582(65)90463-3},
issn = {0029-5582},
journal = {Nuclear Physics},
number = {3},
pages = {658--692},
title = {{The theory of muon capture by complex nuclei: (III). Radiative muon capture}},
url = {http://www.sciencedirect.com/science/article/pii/0029558265904633},
volume = {70},
year = {1965}
}

@article{Fearing:1979be,
      author         = "Fearing, Harold W.",
      title          = "{A Relativistic Calculation of Radiative Muon Capture in
                        Hydrogen and $^{3}$He}",
      journal        = "Phys. Rev.",
      volume         = "C21",
      year           = "1980",
      pages          = "1951-1958",
      doi            = "10.1103/PhysRevC.21.1951",
      reportNumber   = "TRI-PP-79-37",
      SLACcitation   = "%%CITATION = PHRVA,C21,1951;%%"
}

@article{Fearing:1991je,
    author = "Fearing, Harold W. and Welsh, Mark S.",
    title = "{Radiative muon capture in medium heavy nuclei in a relativistic mean field theory model}",
    reportNumber = "TRI-PP-91-75",
    doi = "10.1103/PhysRevC.46.2077",
    journal = "Phys. Rev. C",
    volume = "46",
    pages = "2077--2089",
    year = "1992"
}

@article{Bernard:2000et,
    author = "Bernard, Veronique and Hemmert, Thomas R. and Meissner, Ulf-G.",
    title = "{Ordinary and radiative muon capture on the proton and the pseudoscalar form-factor of the nucleon}",
    eprint = "nucl-th/0001052",
    archivePrefix = "arXiv",
    reportNumber = "FZJ-IKP-TH-2000-02",
    doi = "10.1016/S0375-9474(00)00520-0",
    journal = "Nucl. Phys. A",
    volume = "686",
    pages = "290--316",
    year = "2001"
}

@article{Meissner:1997mq,
    author = "Meissner, T. and Myhrer, F. and Kubodera, K .",
    title = "{Radiative muon capture by a proton in chiral perturbation theory}",
    eprint = "nucl-th/9707019",
    archivePrefix = "arXiv",
    doi = "10.1016/S0370-2693(97)01330-0",
    journal = "Phys. Lett. B",
    volume = "416",
    pages = "36--42",
    year = "1998"
}

@article{Ando:1997es,
    author = "Ando, Shung-ichi and Min, Dong-Pil",
    title = "{Radiative muon capture in heavy baryon chiral perturbation theory}",
    eprint = "hep-ph/9707504",
    archivePrefix = "arXiv",
    reportNumber = "SNUTP-97-105",
    doi = "10.1016/S0370-2693(97)01367-1",
    journal = "Phys. Lett. B",
    volume = "417",
    pages = "177--185",
    year = "1998"
}


%% file: bib/tools.bib
@article{Brun:1997pa,
    author        = "Brun, R. and Rademakers, F.",
    title         = "{ROOT --- An object oriented data analysis framework}",
    journal       = "Nucl. Instrum. Meth. A",
    volume        = "389",
    pages         = "81--86",
    year          = "1997",
    doi           = "10.1016/S0168-9002(97)00048-X"
}

@inproceedings{Verkerke:2003ir,
    author        = "Verkerke, Wouter and Kirkby, David",
    title         = "{The RooFit toolkit for data modeling}",
    eprint        = "physics/0306116",
    archivePrefix = "arXiv",
    pages         = "186--189",
    year          = "2003"
}

@techreport{Skwarnicki:1986xj,
    author      = "Skwarnicki, Tomasz",
    title       = "{A study of the quarkonium states in LISP and Crystal Ball detectors}",
    institution = "DESY, Hamburg",
    reportNumber = "DESY-F31-86-02",
    year        = "1986",
    note        = "Ph.D. Thesis, Institute of Nuclear Physics, Krakow"
}


%% file: bib/used_before.bib
@article{PDG:2025_AluminumProps,
  author        = "Navas, S. and others",
  title         = "{Review of Particle Physics}",
  collaboration = {Particle Data Group Collaboration},
  journal = {Phys. Rev. D},
  volume = {110},
  issue = {3},
  pages = {030001},
  numpages = {5},
  year = {2024},
  month = {Aug},
  publisher = {American Physical Society},
  doi = {10.1103/PhysRevD.110.030001},
  url = {https://link.aps.org/doi/10.1103/PhysRevD.110.030001}
}

@article{Mu2e:2022Run1Sensitivity,
    author        = "{Mu2e Collaboration}",
    title         = "{Mu2e Run I Sensitivity Projections for the Neutrinoless $\mu^- \to e^-$ Conversion Search in Aluminum}",
    journal       = "Universe",
    volume        = "9",
    number        = "1",
    pages         = "54",
    year          = "2023",
    doi           = "10.3390/universe9010054",
    eprint        = "2210.11380",
    archivePrefix = "arXiv",
    primaryClass  = "hep-ex"
}

@article{2018_CALIBBI_SIGNORELLI,
  title = {Charged lepton flavour violation: An experimental and theoretical introduction},
  author = {Calibbi, Lorenzo and Signorelli, Giovanni},
  journal = {La rivista del Nuovo Cimento},
  volume = {41},
  issue = {2},
  pages = {},
  numpages = {},
  year = {2018},
  month = {Feb},
  publisher = {American Physical Society},
  doi = {10.1393/ncr/i2018-10144-0},
  url = {https://www.sif.it/riviste/sif/ncr/econtents/2018/041/02/article/0}
}

@article{2017_CLFV_Szafron_CE_LL,
    author = {Szafron, Robert},
    title = {Radiative Corrections in Bound States},
    doi = "10.5506/APhysPolB.48.2183",
    journal = "Acta Phys. Polon. B",
    volume = "48",
    pages = "2183",
    year = "2017"
}

@misc{2013_CLFV_MU3E_PROPOSAL,
  title         = {Research Proposal for an Experiment to Search for the Decay $\mu \rightarrow eee$},
  author        = {Blondel, A. and {et al}},
  year          = {2013},
  eprint        = {1301.6113},
  archivePrefix = {arXiv},
  primaryClass  = {physics.ins-det},
  url           = {https://arxiv.org/abs/1301.6113}
}

@article{2012_CLFV_Czarnecki,
author="Czarnecki, Andrzej
and Garcia i Tormo, Xavier
and Marciano, William Joseph",
title="Muon decay in orbit spectra for $\mu${\thinspace}-{\thinspace}e conversion experiments",
journal="Hyperfine Interactions",
year="2012",
month="May",
day="01",
volume="210",
number="1",
pages="19--23",
issn="1572-9540",
doi="10.1007/s10751-011-0540-0",
url="https://doi.org/10.1007/s10751-011-0540-0"
}

@article{2008_CLFV_Marciano,
author = {Marciano, William J and Mori, Toshinori and Roney, J. Michael},
issn = {0163-8998},
journal = {Annual review of nuclear and particle science},
language = {eng},
number = {1},
pages = {315-341},
title = {Charged Lepton Flavor Violation Experiments},
volume = {58},
year = {2008},
}

@phdthesis{1997_CLFV_SINDRUM_II_Kaulard_Thesis,
  author       = {J{\"o}rg Q. Kaulard}, 
  school       = {RWTH Aachen University},
  year         = {1997},
  month        = apr,
  url = {https://index.ub.rwth-aachen.de/TouchPointClient_touchpoint/perma.do?q=0%3D%22270101%22+IN+%5B2%5D\&v=sunrise\&b=0\&l=en}
}

@misc{bindingEnergies,
      title={Impact of the nuclear charge distribution on the g-factors and ground state energies of bound muons}, 
      author={Atakan Çavuşoğlu and Bastian Sikora},
      year={2023},
      eprint={2311.16855},
      archivePrefix={arXiv},
      primaryClass={physics.atom-ph},
      url={https://arxiv.org/abs/2311.16855} 
}

@article{Wang_2017,
doi = {10.1088/1674-1137/41/3/030003},
url = {https://doi.org/10.1088/1674-1137/41/3/030003},
year = {2017},
month = {mar},
publisher = {},
volume = {41},
number = {3},
pages = {030003},
author = {Wang, M. and Audi, G. and Kondev, F. G. and Huang, W. J. and Naimi, S. and Xu, Xing},
title = {The AME2016 atomic mass evaluation (II). Tables, graphs and references},
journal = {Chinese Physics C}
}

@article{PlestidAndHill,
  title = {High energy spectrum of internal positrons from radiative muon capture on nuclei},
  author = {Plestid, Ryan and Hill, Richard J.},
  journal = {Phys. Rev. D},
  volume = {103},
  issue = {3},
  pages = {033002},
  numpages = {13},
  year = {2021},
  month = {Feb},
  publisher = {American Physical Society},
  doi = {10.1103/PhysRevD.103.033002},
  url = {https://link.aps.org/doi/10.1103/PhysRevD.103.033002}
}

@techreport{2002_CLFV_SINDRUM_II_Annual_Report,
    author      = {van der Schaaf, Andries},
    title       = {Search for $\mu$-e Conversion with SINDRUM II},
    institution = {Physik-Institut, Universit{\"a}t Z{\"u}rich},
    type        = {Wissenschaftlicher Jahresbericht},
    number      = {April 2001--March 2002},
    pages       = {7--9},
    year        = {2002},
    url         = {https://www.physik.uzh.ch/dam/jcr%3Ae0f9f224-12ed-4ff3-8c05-2bb79b11a081/jb2002.pdf}
}

@article{Watanabe:1993,
title = "Asymmetry and Energy Spectrum of Electrons in Bound-Muon Decay",
journal = "Atomic Data and Nuclear Data Tables",
volume = "54",
number = "1",
pages = "165 - 178",
year = "1993",
issn = "0092-640X",
doi = "https://doi.org/10.1006/adnd.1993.1012",
url = "http://www.sciencedirect.com/science/article/pii/S0092640X83710120",
author = "R. Watanabe and K. Muto and T. Oda and T. Niwa and H. Ohtsubo and R. Morita and M. Morita",
}
